\documentclass[aps,prd,twocolumn,superscriptaddress,longbibliography,nofootinbib]{revtex4-2}

\usepackage[T1]{fontenc}
\usepackage{lmodern}   
\usepackage[utf8]{inputenc}
\usepackage{amsmath,amssymb,amsfonts,bm,mathtools}
\usepackage{graphicx}
\usepackage{array}[=2016-10-06]    

\usepackage{microtype}
\usepackage{hyperref}
\hypersetup{
  pdftitle={Frame covariance of CMBR lensing in the exact 1+3 covariant radiation hierarchy},
  pdfauthor={Tim Gebbie},
  pdfsubject={Cosmic microwave background; 1+3 covariant and gauge-invariant cosmology},
  pdfkeywords={cosmic microwave background, relativistic cosmology, 1+3 covariant formalism, Boltzmann hierarchy, gravitational lensing}}
\hypersetup{colorlinks=true,linkcolor=blue,citecolor=blue,urlcolor=blue}

\def\D{\mathrm{D}}                  
\def\dd{\mathrm{d}}
\def\la{\langle}                    
\def\ra{\rangle}
\def\ep{\varepsilon}
\def\ts{\textstyle}
\def\st{\sigma_{\!_T}}              
\def\ne{n_{\!_E}}
\def\vb{v_{\!_B}}

\begin{document}

\title[Frame covariance of CMBR lensing]{Frame covariance of CMBR lensing in the
exact $1+3$ covariant radiation hierarchy}

\author{Tim Gebbie}
\email{tim.gebbie@uct.ac.za}
\affiliation{Department of Statistical Sciences, University of Cape Town,
Rondebosch 7701, South Africa}

\date{\today}

\begin{abstract}
The exact $1+3$ covariant Boltzmann hierarchy for the cosmic microwave background contains acceleration, shear and vorticity couplings whose coefficients grow linearly in $\ell$. We show that all are the multipole form of the aberration operator in the photon Liouville equation. Under a change of threading, this operator shifts by the along-ray generator of a Lorentz boost. The acceleration coupling of $\ell$ to $\ell\pm1$ and the shear coupling of $\ell$ to $\ell\pm2$ are therefore frame-covariant descriptions of the same operator, related by an algebraic, pointwise relabelling of direction. With emitter and observer frames fixed, the observed sky is independent of the intermediate threading at every $\ell$. We derive this in $1+3$ covariant and gauge-invariant variables without constructing second-order gauge-invariant variables. The $O(\ep^2\ell)$ content is gravitational lensing, as known in metric variables. Linearisation about a Friedmann--Lema\^{i}tre--Robertson--Walker background removes these couplings, requiring lensing to be reinstated from the geodesic equation. Keeping the threading arbitrary makes frame covariance explicit and places observer aberration in the same operator. In the Newtonian threading we compute both halves of the deflection: the equivalence-principle contribution enters through aberration, while the spatial-curvature contribution arises from free streaming on perturbed spatial sections and reduces to the same operator. The lensing efficiency follows from the direction derivative. The angular-power-spectrum contribution is quartic in the potential: for a deflection statistically independent of the source, the two fourth-order terms carrying $\ell^2$ cancel where the spectrum is smooth on the deflection scale, and lensing redistributes power without adding ensemble variance.
\end{abstract}


\maketitle

\section{Introduction}
\label{sec:intro}

Cosmic microwave background radiation (CMBR) anisotropies are almost universally
computed by
linearising the Einstein--Boltzmann system about a Friedmann--Lema\^{i}tre--Robertson--Walker
(FLRW) background. The procedure is enormously successful, but it is a
\emph{truncation}: terms present in the exact equations are removed before the
hierarchy is solved, and they cannot be recovered from the linearised hierarchy
or from its solution alone. They can be restored only by deriving the relevant
higher-order hierarchy, or by supplying the equivalent geodesic remapping.

The $1+3$ covariant and gauge-invariant (CGI) formulation \cite{Ellis1971,
EllisVanElst1998, MaartensGebbieEllis1999} makes this explicit. Written
covariantly, the radiation multipole divergence equations contain couplings
between the brightness multipoles and the kinematic quantities of the threading
congruence --- the shear $\sigma_{ab}$, the acceleration $A_a$ and the vorticity
$\omega^a$ --- with coefficients that grow \emph{linearly in $\ell$}. The key
advantage of the $1+3$ Lagrangian approach for this class of calculation is that
it is not encumbered by the construction of higher-order gauge-invariant
variables \cite{BruniMatarreseMollerachSonego1997}: the variables are
gauge-invariant by the Stewart--Walker lemma from the outset, because they vanish
in the background.

The physical question is what this linear-in-$\ell$ growth does to the standard
scaling argument, which places the onset of gravitationally nonlinear
corrections at $\ell \gtrsim 5000$ \cite{Seljak1996, MartinezGonzalezSanzSilk1994}.
That argument proceeds from the Poisson relation between potentials and
densities and a cancellation factor $(k\,\delta\eta)^{-1}$, and concludes that
the Rees--Sciama (RS) source scales as $k^{2}P(k)$ and is negligible. If the
exact hierarchy supplies an additional factor of $\ell$ at the level of the
source, the conclusion does not follow without further argument.

The couplings were identified in this form, and the question posed, in
Refs.~\cite{Gebbie2000, GebbieThesis1999} and the replacement version of the
first \cite{Gebbie2026corr}.
In metric variables the answer is known: the second-order Boltzmann hierarchy
contains the corresponding couplings, with coefficients growing like $\ell$, and
identifies them as weak lensing \cite{BenekeFidler2010}.

This paper derives that answer inside the $1+3$ hierarchy, in one formalism. 
Every $O(\ell)$ coupling of the exact hierarchy is the
multipole form of one operator, the aberration term of the photon Liouville
equation, which rotates the direction label of the threading relative to the
photon. Its content at $O(\ep^2\ell)$ is gravitational lensing of the CMBR, together
with the observer's kinematic aberration, and nothing else.
The extra factor of $\ell$ is therefore real, and it is lensing. It leaves the
scaling argument for the Rees--Sciama term untouched, and its contribution to the
angular power spectrum is quartic in the potential, where the linear spectrum is
quadratic. At that order the two terms that carry $\ell^2$ cancel where the spectrum is
smooth on the deflection scale, so an $O(\ell)$ source does not become
$O(\ell^2)$ growth in the power: for a deflection statistically independent of
the source, lensing moves power between multipoles and adds no ensemble
variance.

Those two statements belong together. The nonlinear source \emph{matters}: it is
a derivation, inside general relativity and inside the hierarchy that should
contain it, of an effect otherwise obtained from outside; and it adds nothing to
the power, its contribution to $C_\ell$ being quartic in the potential and equal
to standard lensing. Either on its own invites the other to be rediscovered as a
new effect.

Lensing is not normally obtained this way. The linear treatment removes the
couplings, and so re-introduces lensing afterwards, from the geodesic equation,
as a remapping of a separately computed unlensed sky \cite{LewisChallinor2006}.
Second-order codes keep the couplings up to recombination but exclude them from
the late-time sources and defer to the same remapping \cite{Pettinari2013}. In
the $1+3$ hierarchy the couplings are never removed. Lensing is present as an
exact kinematic coupling, in variables that need no second-order gauge
construction, and the two halves of the light bending enter through different
terms of the hierarchy: the equivalence-principle half through the aberration
operator, the spatial-curvature half through free streaming on the threading's
perturbed spatial sections. The lensing efficiency $(\chi_*-\chi)/\chi_*$ comes
out of the distance factor carried by the direction derivative of the
free-streamed field.

What is claimed here is the frame. The identification of these couplings as
lensing, and their growth in $\ell$, are known in metric variables
\cite{BenekeFidler2010}; the paper cites that work for them. But a metric
treatment that places the observer at rest has chosen a threading that is
shear-free for scalar modes, and so carries no $\ell\pm2$ shear coupling in the
sector it computes --- there is nothing in it whose frame dependence could be
examined. Because the $1+3$ threading is a
choice made inside the equations rather than before they are written, the frame
covariance of the content can be exhibited. We are not aware of a treatment of
these couplings that does so, or that places the observer's own aberration in
the same operator as the lensing.

Section~\ref{sec:conventions} fixes the conventions: the signature and
projector, the scalar harmonics and the mode normalisation, the PSTF contraction
identities and $\beta_\ell$, and what it means to fix the threading
inside the equations. Section~\ref{sec:hierarchy} constructs the exact
hierarchy from the Liouville equation, separates the redshift and aberration
terms, and fixes the sign of the $\ell-2$ shear coupling.
Section~\ref{sec:weaklynl} isolates the $O(\ep^2\ell)$ system that linearisation
discards and reduces its four couplings to PSTF form.
Section~\ref{sec:content} identifies that content: the aberration operator
carries the equivalence-principle half of the deflection, free streaming on the
perturbed sections carries the spatial-curvature half, the lensing efficiency
comes out of the distance factor in the direction derivative, and an enumeration
shows nothing is left over. Section~\ref{sec:frame} establishes frame covariance by
computing the division in the Newtonian and CDM threadings separately and
exhibiting the boost that relates them. Section~\ref{sec:quartic} shows the
contribution to the angular power spectrum is quartic.
Section~\ref{sec:discussion} places the result against the second-order Boltzmann
literature.

\section{Conventions and the $1+3$ kinematic frame}
\label{sec:conventions}

Every sign and every factor of $\ell$ in what follows is produced by three
choices: the normalisation of the scalar harmonics, the metric signature, which
enters once per contracted index and so distinguishes a rank-raising coupling
from a rank-lowering one, and the weights used in removing traces. None of the
three is visible in a finished expression. A calculation carrying an
inconsistent choice is still dimensionally consistent and still has the right
limits; it simply has the wrong sign somewhere, and every later
line agrees with it.

This section fixes the conventions before anything is derived from them, and
every subsequent sign and $\ell$-weight cites the line of
Table~\ref{tab:conventions} it rests on. The conventions are those of the
covariant CMBR papers that introduced this formalism \cite{GebbieEllis1998,
GebbieDunsbyEllis1999}, and the kinematic variables are those of the $1+3$
literature more broadly \cite{Ellis1971, EllisVanElst1998}.

\subsection{Signature, the projector and the kinematic decomposition}
\label{sec:conv-kin}

The signature is $(-,+,+,+)$ and $u^au_a=-1$ [Table~\ref{tab:conventions},
C1.1].\footnote{%
Reference~\cite{GebbieEllis1998} does not state a signature, but fixes one: it
writes $e^ae_a=1$, $e^au_a=0$ for the direction on the sphere, and defines
$p_{ab}=h_{ab}-e_ae_b$ with $p_a{}^a=2$, a trace that holds only for
$e^ae_a=+1$. The conventions of that work and of
Ref.~\cite{MaartensGebbieEllis1999} therefore agree, which matters here because
the harmonic normalisation is taken from the first and the signature from the
second.}
The tensor $h_{ab}=g_{ab}+u_au_b$ projects into the rest space of $u^a$ [C1.2];
in this signature it is \emph{positive} semi-definite there, so contracting a
pair of spatial indices introduces no sign. That is worth stating, because the
sign a contraction carries is one of the two places a relative sign between a
rank-raising and a rank-lowering coupling can come from, and in this signature
it is not the one that does [C1.6].

The projected derivative is $\D_c = h_c{}^d\nabla_d$ acting on an already
projected tensor [C2.1], and the kinematic decomposition is
\begin{equation}
  \nabla_b u_a = -A_a u_b + \ts\frac13 h_{ab}\Theta
  + \ep_{abc}\omega^c + \sigma_{ab} ,
  \label{eq:kinematics}
\end{equation}
with $\Theta=\D^au_a$, $A_a=\dot u_a$, $\sigma_{ab}=\D_{\la a}u_{b\ra}$ and
$\omega_a=-\frac12\mathrm{curl}\,u_a$ [C2.2, C2.3]. Contracting
Eq.~(\ref{eq:kinematics}) with $u^b$ returns $+A_a$, which fixes the sign of the
acceleration term. Overdots denote proper time along $u^a$; primes denote
conformal time $\eta$, with $\dd t = a\,\dd\eta$. Directions satisfy
$e^ae_a=+1$, $e^au_a=0$, and a photon has $p^a=E(u^a+e^a)$ [C1.5].

\subsection{Orders, and what they count}
\label{sec:orders}

Three relations are used throughout, in the sense fixed by
Ref.~\cite{GebbieDunsbyEllis1999}: $A=B$ holds in the exact theory, $A\simeq B$
holds in the almost-Friedmann--Lema\^itre sense, and $A\approx B$ holds in
relation to some further smallness parameter, such as the Thomson relaxation
time or the ratio of radiation to matter density
[Table~\ref{tab:conventions}, C0.1].

The counting behind $O(\ep^2\ell)$ is of the same kind, and it is not a
perturbation series. A quantity that vanishes in the exact
Friedmann--Lema\^itre limit is $O(\ep)$ --- gauge-invariant for that reason, by
the Stewart--Walker lemma --- once suitably normalised, as in
$\sqrt{\sigma_{ab}\sigma^{ab}}/H<\ep$. A product of two such quantities is
$O(\ep^2)$. The second label counts powers of the multipole rank, which enters
through Eq.~(\ref{eq:dPide}) and not through any expansion in amplitude. So
$O(\ep^2\ell)$ names a magnitude: a product of two almost-Friedmann--Lema\^itre
quantities carrying one factor of the rank. That such terms are kept while
$O(\ep^2\ell^0)$ terms are not is a statement about physical scales and about
what is measurable, not about where a series was truncated.

Nothing here constructs a perturbation series in a gauge. What this paper adds
to the almost-Friedmann--Lema\^itre treatment of
Ref.~\cite{GebbieDunsbyEllis1999} is called, as it is called there, a nonlinear
correction; the term \emph{second order} is used below only for the metric
treatments to which it belongs. A reader coming from
Ref.~\cite{Gebbie2000} should note that it uses \emph{second order} in the
looser sense, for the same $O(\ep^2)$ content.

\subsection{Scalar harmonics and the mode normalisation}
\label{sec:conv-harmonics}

The scalar harmonics satisfy the Helmholtz equation
$\D^a\D_a Q = -k^2 Q$ with $k$ the physical wavenumber and $Q$ independent of
time [C3.1], and the PSTF harmonic tensors are
\begin{equation}
  Q_{A_\ell} = (-k)^{-\ell}\,\D_{\la A_\ell\ra} Q .
  \label{eq:harmonic}
\end{equation}
The eigenfunctions $Q$ are real and the stripped factor $(-k)^\ell$ is real, so
every $Q_{A_\ell}$ is real and carries no complex phase [C3.2]. This is a
property of the representation. Contracting already-defined spatial PSTF tensors
with the positive rest-space metric introduces no additional metric-signature or
complex-phase factor, but the signs generated by spatial derivatives, by the
Helmholtz equation and by the rank-dependent definition (\ref{eq:harmonic})
remain and are tracked separately: reality does not make a change of harmonic
rank through derivatives sign-free. Equation
(\ref{eq:harmonic}) at $\ell=1$ gives $\D_aQ = -k\,Q_a$, and iterating it gives
the two relations that every $\ell$-weight below rests on --- a PSTF gradient
raises rank with a factor $-k$,
\begin{equation}
  \D_{\la a}Q_{A_\ell\ra} = -k\,Q_{aA_\ell} ,
  \label{eq:raise}
\end{equation}
while a divergence lowers it with a factor $+k$,
\begin{equation}
  \D^aQ_{aA_\ell} = +\frac{\ell+1}{2\ell+1}\,k\,Q_{A_\ell}
  \label{eq:lower}
\end{equation}
in the flat case [C4.1, C4.2, C4.4]. The opposition of those two signs is not a
convention: it follows from Eq.~(\ref{eq:harmonic}) and the Helmholtz equation,
and it is why free streaming couples $\ell-1$ and $\ell+1$ with opposite signs
in mode form although both of its covariant couplings carry a plus [C4.5].

A scalar is expanded as $X = -\sum_k X(k)\,Q$. The minus is a stipulation, not a
derivation --- no source expands a scalar --- but it is forced by consistency
with the conventions adopted here, in which
\begin{equation}
  \D_aX = +k\,X(k)\,Q_a ,
  \qquad
  \D_{\la a}\D_{b\ra}X = -k^2X(k)\,Q_{ab} ,
  \label{eq:scalarmode}
\end{equation}
and under Eq.~(\ref{eq:harmonic}) those two hold together only with that sign
[C3.5]. The wavenumber is physical throughout [C3.1], so it carries the time
dependence, $k\propto1/a$ and so $\dot k=-Hk$, and no scale factor stands
beside it in Eqs.~(\ref{eq:raise})--(\ref{eq:scalarmode}); a comoving label
$q=ak$ would put that dependence into the coefficients instead. The harmonics
themselves are time-independent: each $\D$ in Eq.~(\ref{eq:harmonic}) carries a
factor $1/a$ and each $k^{-1}$ carries an $a$, so $Q_{A_\ell}$ does not evolve.
Any time derivative of a mode amplitude below is to be read with $\dot k=-Hk$;
that is what a scale factor written beside $k$ would otherwise hide. Multipoles are expanded as $\tau_{A_\ell} = \sum_k \tau_\ell(k)Q_{A_\ell}$
[C3.4]; because the mode recursion that follows from
Eqs.~(\ref{eq:raise})--(\ref{eq:lower}) has real coefficients, the $\tau_\ell(k)$
are real and carry no $\ell$-dependent phase of their own.\footnote{%
Reference~\cite{GebbieEllis1998} prints two possibilities. We state
which is followed. Its Eq.~(48) is Eq.~(\ref{eq:harmonic}) above and has a real
stripped factor; its Eq.~(59) prints $(-1)^\ell$, which is the phase of
$Q_{A_\ell}=(+ik)^{-\ell}\D_{\la A_\ell\ra}Q$, a different definition.
\emph{This paper follows Eq.~(48).} Nothing below rests on the choice, because
the couplings are settled in index algebra where no harmonic appears at all;
the working is in the supplementary material \cite{Bundle}.}

That recursion is, by a change of normalisation and nothing else, the
free-streaming equation of the standard treatments. It is Eq.~(7) of
Ref.~\cite{WilsonSilk1981} for $\ell>2$, and Eq.~(8) of its sequel
Ref.~\cite{Wilson1983}, which carries the spatial-curvature factor along; it is
Eq.~(6) of Ref.~\cite{HuSugiyama1995PRD}; and on replacing $\beta_\ell$ by
$\alpha_\ell^{-1}(2\ell+1)$ it is Eqs.~(63) and~(64) of
Ref.~\cite{MaBertschinger1995}, identical in the synchronous and the conformal
Newtonian gauge, and Eq.~(3d) of Ref.~\cite{SeljakZaldarriaga1996}. What the
five matches check is the $\ell$-weights. They cannot also settle the phase
convention, because each of those papers absorbs its own $\lambda^\ell$ into its
multipole definition before printing, and they do not all absorb the same one:
the stripped factor is real in Ref.~\cite{HuSugiyama1995PRD}, whose angular
functions are built on $(-k)^{-\ell}$ as Eq.~(\ref{eq:harmonic}) is and reduce
to $(-i)^\ell e^{i\mathbf{k}\cdot\mathbf{x}}P_\ell$ in the flat-space limit,
and imaginary in Eq.~(1) of Ref.~\cite{Wilson1983}, which strips $(ik)^{-\ell}$.
The two differ by $\lambda^\ell$ at $\lambda=i$, which is why the same-sign
bracket under an overall $-i$ and the opposite-sign real bracket are one
equation. Reference~\cite{Pitrou2009} writes both decompositions and prints the
factor between them, $N_\ell=i^\ell\sqrt{(2\ell+1)/4\pi}$, at its Eq.~(8.4).
The working is in the supplementary material \cite{Bundle}.

\subsection{PSTF algebra}
\label{sec:conv-pstf}

The direction basis is the PSTF product $O^{A_\ell}=e^{\la A_\ell\ra}$, and
$\tau(x,e)=\sum_\ell \tau_{A_\ell}O^{A_\ell}$ [C5.1]. The normalisation
$\beta_\ell = O^{A_\ell}O_{A_\ell} = \ell!/(2\ell-1)!!$ [C5.2] satisfies
\begin{equation}
  \frac{\beta_\ell}{\beta_{\ell-1}} = \frac{\ell}{2\ell-1} ,
  \qquad
  \frac{\beta_\ell}{\beta_{\ell+1}} = \frac{2\ell+1}{\ell+1}
  \label{eq:betaratios}
\end{equation}
[C5.3]. The contraction identities used throughout are
\begin{equation}
\begin{gathered}
  e_a O^{aA_\ell} = \frac{\ell+1}{2\ell+1}\,O^{A_\ell} ,
  \\
  e_ae_b O^{abA_\ell}
  = \frac{(\ell+1)(\ell+2)}{(2\ell+1)(2\ell+3)}\,O^{A_\ell} ,
\end{gathered}
  \label{eq:identities}
\end{equation}
with limits $\frac12$ and $\frac14$ at high $\ell$ [C6.1, C6.2]. Both weights
are real and positive, and neither carries a phase. The same numbers
$(\ell+1)/(2\ell+1)$ appear in Eq.~(\ref{eq:lower}) for an unrelated reason ---
there as a divergence of a gradient in space, here as a contraction of $e$ with
$O$ on the sphere --- and the two should not be run together [C6.7].

\subsection{Threading as a choice inside the equations}
\label{sec:threading}

The choice of $u^a$ is a choice of \emph{threading}, not of a coordinate gauge.
Let $\tilde u^a\simeq u^a+v^a$, $v_au^a=0$. To first order
\cite{MaartensGebbieEllis1999}
\begin{equation}
\begin{gathered}
  \tilde A_a \simeq A_a + \dot v_a + Hv_a ,
  \qquad
  \tilde\sigma_{ab} \simeq \sigma_{ab} + \D_{\la a}v_{b\ra} ,
  \\
  \tilde\omega_a \simeq \omega_a - \ts\frac12\,{\rm curl}\,v_a ,
\end{gathered}
  \label{eq:frame-transf}
\end{equation}
and the shear, vorticity and trace parts of $\D_bv_a$ recombine in $K_\perp^a$:
\begin{equation}
  \tilde K_\perp^a - K_\perp^a \simeq
  \Bigl(\frac{\dd v^a}{\dd t} + Hv^a\Bigr)_{\!\perp} ,
  \label{eq:dK}
\end{equation}
where $\dd/\dd t = u^a\nabla_a + e^a\D_a$ is the derivative along the ray. This
is the rate at which the event-by-event aberration map
$\tilde e^a = e^a - v_\perp^a$ between the two frames changes along the ray. The
two hierarchies are one equation in two coordinate systems on photon phase
space, and their solutions differ at each event by Doppler plus aberration. With
the frames at the endpoints fixed, the observed sky does not depend on the
threading in between. Equation~(\ref{eq:dK}) is the direction counterpart of
the total-derivative shift of the redshift integrand in
Ref.~\cite{ZibinScott2008}.

Two threadings are used below. In the Newtonian threading the shear vanishes
for scalar modes, $\tilde\sigma_{ab}=0$, and $\tilde A_a\simeq\D_a\Phi_A$;
App.~\ref{app:tensor} says what changes when tensors are present. In the
CDM threading $u^a_{\!_C}=\tilde u^a+v^a_{\!_C}$ is comoving with the cold dark
matter; that matter is pressureless and collisionless, so its worldlines are
geodesic and the acceleration vanishes. Equation~(\ref{eq:frame-transf}) then
gives
\begin{equation}
  \sigma^{\!_C}_{ab} = \D_{\la a}v^{\!_C}_{b\ra} ,
  \qquad
  \tilde A_a = -\bigl(\dot v^{\!_C}_a + Hv^{\!_C}_a\bigr) .
  \label{eq:twoframes}
\end{equation}
The whole CDM-frame shear is the gradient of the boost from the Newtonian
threading, and the whole Newtonian acceleration is its rate of change. Neither
is suppressed relative to the other: for a sub-horizon mode
$v\sim(H/k)\Delta$ is small, but $e^b\D_bv\sim kv\sim H\Delta\sim\sigma$ is not.

\subsection{The free field and where the frame enters it}
\label{sec:weyl}

The Ricci curvature is fixed pointwise by the matter through the field
equations; the free gravitational field is the Weyl tensor, and relative to
$u^a$ it splits into an electric and a magnetic part,
\begin{equation}
  E_{ab}=C_{acbd}u^cu^d ,
  \qquad
  H_{ab}=\ts\frac12\,\ep_{acd}\,C^{cd}{}_{be}\,u^e ,
  \label{eq:weyl}
\end{equation}
both PSTF [Table~\ref{tab:conventions}, C2.3]. In this split the Bianchi
identities become a set of evolution and constraint equations of exactly
Maxwell's form, with the matter as source
\cite{Ellis1971,EllisVanElst1998,VanElstEllis1998}; they are not reproduced
here. Three consequences are used below, and one further relation of the same
family --- the Gauss relation for a shear-free irrotational congruence --- is
what Sec.~\ref{sec:curvhalf} needs to fix its conformal factor.

First, the tidal field never enters the Boltzmann hierarchy directly. It enters
through the shear, because the shear propagation equation carries $-E_{ab}$ as
its source, and it is $\sigma_{ab}$ --- not $E_{ab}$ --- that
Table~\ref{tab:coefficients} couples to the multipoles. The lensing half of
Sec.~\ref{sec:curvhalf} is in that sense an $E_{ab}$ statement made through
$\sigma_{ab}$, while the aberration half of Sec.~\ref{sec:lensing} is a
statement about $A_a$ and $\sigma_{ab}$ as properties of the chosen congruence.
The two halves are already distinguishable at the level of the field equations,
before any photon is propagated.

Second, $E_{ab}$ and $H_{ab}$ are defined relative to $u^a$ and mix into each
other under a change of threading, as the electric and magnetic parts of the
Maxwell field do under a boost. The Weyl tensor is the frame-independent
object; ``the tidal field'' is not. A claim that the deflection is produced by
a tidal field is therefore a claim made in a threading, and the division of
Sec.~\ref{sec:frame} is the photon-side image of this one.

Third, $H_{ab}$ has no Newtonian analogue, and its part here is structural
rather than that of a further leading scalar source. That $H_{ab}=0$ for
first-order scalar modes holds relative to the congruence whose rest spaces and
projected derivative define that decomposition: under a change of physical
threading the electric and magnetic projections of one Weyl tensor are
repartitioned, as the acceleration, shear, vorticity and photon direction are.
Counting the relative velocity as $O(\ep)$ about one fixed FLRW congruence makes
the boost-induced magnetic part $O(\ep^2)$ in that bookkeeping, which neither
makes the electric--magnetic split observer-independent nor privileges the
original congruence. The leading angular transport term is displayed through
$A_a$, $\sigma_{ab}$, $\omega_a$ and $K^a_\perp$ directly; the Weyl mixing is
the curvature-level reason that two threadings can encode one optical deflection
differently, and a purely Newtonian formulation has nowhere to store the
magnetic part.\footnote{%
A further consequence is not used here but bounds how far the formalism can be
pushed. The density in the $1+3$ equations is the density measured by $u^a$,
and spatial averaging on the surfaces orthogonal to $u^a$ does not commute with
the evolution, so averaged equations acquire backreaction terms
\cite{ClarksonEllisLarenaUmeh2011, Korzynski2015}. Nothing at $O(\ep^2\ell)$ here depends on an
averaging operation: the hierarchy is written at a point and the frame shifts
of Sec.~\ref{sec:frame} are exact maps between congruences, not averages. It
becomes a live question once the frame is fixed at higher order, or outside
$\Lambda$CDM, where the background against which $\ep$ is counted is itself an
averaged quantity.}

Fourth, the tensor sector needs no separate treatment. At first order the scalar
sector carries no magnetic part; gravitational waves carry both parts, and they
reach the photons the way everything else does --- through $\sigma_{ab}$ in
$K^a$, Eq.~(\ref{eq:K}). What they produce are the couplings of
Table~\ref{tab:coefficients}, with the same $\ell$-weights and the same single
operator, so the identification of Sec.~\ref{sec:content} and the frame
covariance of Sec.~\ref{sec:frame} cover them unchanged. What is not covered is
the amplitude, which is set by the tensor shear rather than by
$\D_{\la a}\D_{b\ra}\Phi_A$ and is not computed here. In a metric treatment the
tensor contribution is a decomposition made before the equations are written;
here it is already in $\sigma_{ab}$. Appendix~\ref{app:tensor} sets this out,
and says why this is important: the Newtonian threading is shear-free for scalar
modes,
and not when tensors are present.


\section{The exact covariant Boltzmann hierarchy}
\label{sec:hierarchy}

The $1+3$ formulation carries the radiation anisotropy as a set of
projected, symmetric, trace-free (PSTF) multipoles of the brightness, defined
with respect to a timelike congruence $u^a$ that is chosen rather than fixed by
a gauge condition. The Liouville equation for the photon distribution then
becomes an infinite set of coupled divergence equations for those multipoles.
Nothing in that construction requires the spacetime to be close to FLRW: the
hierarchy is exact in the kinematics and in the anisotropy, and the only
approximation made anywhere in this section is in the Thomson collision term.

The structure that matters for what follows is already visible in the way the
Liouville operator acts. Transporting the photon distribution along the ray
changes it in two distinct ways: the photon energy changes, and the direction
label carried by the threading rotates relative to the photon. The first is a
redshift term and the second an aberration term, and in multipole form the two
are not interchangeable --- the aberration term differentiates the brightness
with respect to direction, so acting on a pattern with angular scale $1/\ell$ it
returns a factor of $\ell$ that the redshift term does not. Separating them is
what makes the $\ell$-growth of the kinematic couplings a statement about one
operator rather than a property of a list of coefficients, and
Table~\ref{tab:coefficients} records the split for every coupling in the
hierarchy.

This section builds that structure: the brightness multipoles and the Liouville
operator, the exact divergence hierarchy in both its energy-resolved and
integrated forms, the redshift--aberration split with its reduction shown, and
the sign of the $\ell-2$ shear coupling.
\label{sec:formulation}

\subsection{Brightness multipoles and the Liouville operator}
\label{sec:multipoles}

The brightness multipoles are the energy moments of the distribution function,
$\Pi(x,e) = \int E^3 f\,\dd E = \sum_\ell \Pi_{A_\ell}O^{A_\ell}$, on the
projected symmetric trace-free (PSTF) basis $O^{A_\ell} = e^{\la A_\ell\ra}$
[Table~\ref{tab:conventions}, C5.1; Ref.~\cite{GebbieEllis1998} Eq.~(15), the
expansion being its Eq.~(16)]. The temperature anisotropy is
defined covariantly by $T(x,e) = T(x)\,[1 + \tau(x,e)]$,
\begin{equation}
  \tau(x,e) = \sum_{\ell\ge 1} \tau_{A_\ell}(x)\, O^{A_\ell} ,
  \label{eq:multipole}
\end{equation}
which is exact, $T(x)$ being fixed as the monopole of $T(x,e)$. Inverting
Eq.~(\ref{eq:multipole}) against the PSTF basis recovers $\tau_{A_\ell}$
[Ref.~\cite{MaartensGebbieEllis1999} Eq.~(79)]. What relates the two sets of
multipoles is the energy integration, and it is short enough to give.

The radiation stress tensor is the second moment of the distribution on the
future null cone, $T^{ab}_{\!_R}=\int p^ap^bf\,\dd^3p$ with
$\dd^3p=E\,\dd E\,\dd\Omega$ the covariant volume element there
[Ref.~\cite{MaartensGebbieEllis1999} Eq.~(46)]. With $p^a=E(u^a+e^a)$ [C1.5]
the energy integral in each component is the one already performed in $\Pi$,
so the dynamic quantities of the radiation are angular moments of the
brightness alone,
\begin{equation}
  \rho_{\!_R} = \oint\Pi\,\dd\Omega , \quad
  q^{\!_R}_a = \oint\Pi\,O_a\,\dd\Omega , \quad
  \pi^{\!_R}_{ab} = \oint\Pi\,O_{ab}\,\dd\Omega ,
  \label{eq:moments}
\end{equation}
which by [C5.5] are $\Delta_\ell\Pi_{A_\ell}$ with
$\Delta_\ell=4\pi\beta_\ell/(2\ell+1)$: the three coefficients are $4\pi$,
$\frac{4\pi}{3}$ and $\frac{8\pi}{15}$, which is Eq.~(49) of
Ref.~\cite{MaartensGebbieEllis1999}. Only the monopole survives an angular
average, so $\rho_{\!_R}=4\pi\Pi_{\varnothing}$, writing $\Pi_\varnothing$ for
the rank-zero brightness.

The temperature in a direction is defined bolometrically: $T(x,e)$ is the
temperature of the blackbody carrying the brightness seen in that direction.
Taking the ratio to the all-sky bolometric temperature $T_{\!_R}$, so that the
radiation constant cancels between the two,
\begin{equation}
  \Bigl[\frac{T(x,e)}{T_{\!_R}(x)}\Bigr]^{4}
  = \frac{\Pi(x,e)}{\Pi_{\varnothing}}
  = 1 + \frac{4\pi}{\rho_{\!_R}}\sum_{\ell\ge1}\Pi_{A_\ell}O^{A_\ell} ,
  \label{eq:bolometric}
\end{equation}
which is exact and is Eqs.~(77) and~(78) of
Ref.~\cite{MaartensGebbieEllis1999}. Taking the fourth root and keeping the
term linear in the anisotropy, $(1+x)^{1/4}=1+\frac14x+O(x^2)$,
\begin{equation}
  \tau_{A_\ell} \simeq \frac{\pi}{\rho_{\!_R}}\,\Pi_{A_\ell}
  \qquad (\ell\ge1) ,
  \label{eq:taupi}
\end{equation}
its Eq.~(80); the two lowest cases are
$\tau_a\simeq 3q^{\!_R}_a/4\rho_{\!_R}$ and
$\tau_{ab}\simeq 15\pi^{\!_R}_{ab}/8\rho_{\!_R}$. What is dropped is quadratic
in the anisotropy, and that is also the order at which the two normalisations
of the mean temperature part company: the monopole of $T(x,e)$, which makes
Eq.~(\ref{eq:multipole}) exact, and the all-sky bolometric $T_{\!_R}$ used
here. Equation~(\ref{eq:taupi}) therefore holds in the
almost-Friedmann--Lema\^itre sense [C0.1] and not beyond it. Nothing below
needs more: the hierarchy is carried in $\Pi_{A_\ell}$ throughout, and
Eq.~(\ref{eq:taupi}) is used only to read its multipoles as a temperature.

\subsection{The exact divergence hierarchy}
\label{sec:divergence}

For $\ell>3$ the brightness multipoles obey \cite{MaartensGebbieEllis1999}
\begin{widetext}
\begin{equation}
\begin{split}
  &\dot\Pi^{\la A_\ell\ra} + \ts\frac43\Theta\,\Pi^{A_\ell}
  + \D^{\la a_\ell}\Pi^{A_{\ell-1}\ra}
  + \frac{(\ell+1)}{(2\ell+3)}\D_b\Pi^{bA_\ell}
  - \frac{(\ell+1)(\ell-2)}{(2\ell+3)}A_b\Pi^{bA_\ell}
  + (\ell+3)A^{\la a_\ell}\Pi^{A_{\ell-1}\ra}
  - \ell\,\omega^b\ep_{bc}{}^{\la a_\ell}\Pi^{A_{\ell-1}\ra c} \\
  &\quad - \frac{(\ell-1)(\ell+1)(\ell+2)}{(2\ell+3)(2\ell+5)}
    \sigma_{bc}\Pi^{bcA_\ell}
  + \frac{5\ell}{(2\ell+3)}\sigma_b{}^{\la a_\ell}\Pi^{A_{\ell-1}\ra b}
  + (\ell+2)\sigma^{\la a_\ell a_{\ell-1}}\Pi^{A_{\ell-2}\ra} \\
  &= -\ne\st\left[\Pi^{A_\ell} - \Pi^{\la A_{\ell-1}}\vb^{a_\ell\ra}
    - \left(\frac{\ell+1}{2\ell+3}\right)\Pi^{A_\ell a}v_{\!_Ba}\right]
    + O(\ep\vb^2,\vb^3) .
\end{split}
  \label{eq:hierarchy-exact}
\end{equation}
\end{widetext}
This is exact in the kinematics and in the anisotropy; the only truncation is
in the Thomson term, and it is written there. The $\ell-2$ shear coupling is
$+(\ell+2)$.\footnote{%
Equation~(71) of Ref.~\cite{MaartensGebbieEllis1999} prints $-(\ell+2)$, as do
its Eqs.~(88) and (90). This is a misprint. Two independent routes give
$+(\ell+2)$ at general $\ell$: the energy integration of that paper's own
Eq.~(70), and the reduction of the Liouville operator in
Sec.~\ref{sec:liouville}, which is carried out there without reference to this
equation and reproduces every other coefficient in it. Two further arguments
agree at $\ell=2$; those three routes are set out in the supplementary
material \cite{Bundle}. Equation numbers are those of the published article, whose
numbering is that of the preprint.}
Every gravitational coupling sits on the left-hand side.

Three couplings grow as $\ell$: the shear couples $\ell$ to $\ell\pm2$, the
acceleration $\ell$ to $\ell\pm1$, and the vorticity $\ell$ to itself
(Table~\ref{tab:coefficients}). At linear order about FLRW every one of them is
a product of two first-order quantities and is dropped. In what follows they
are kept.

\subsection{Redshift and aberration}
\label{sec:liouville}

The origin of the $O(\ell)$ growth is visible one step earlier, in the
Liouville equation from which Eq.~(\ref{eq:hierarchy-exact}) is reduced. Since
$p^b\nabla_bp^a=0$, and using Eq.~(\ref{eq:kinematics}),
\begin{align}
  \frac{\dd E}{\dd\lambda} &= -E^2\bigl(\ts\frac13\Theta + A_ae^a
    + \sigma_{ab}e^ae^b\bigr) ,
  \label{eq:dE}\\
  \frac{\dd e^{\la a\ra}}{\dd\lambda} &= -E\,K_\perp^a ,
  \label{eq:de}
\end{align}
with
\begin{equation}
  K^a \equiv A^a + \sigma^a{}_be^b + \ep^a{}_{bc}e^b\omega^c ,
  \qquad
  K_\perp^a \equiv (h^a{}_b - e^ae_b)K^b .
  \label{eq:K}
\end{equation}
Both rates come from one contraction, $p^b\nabla_bu^a$. Its component along
$e^a$ is the change of energy measured by the threading; its transverse
component is the change of direction measured by the threading. Integrating the
Liouville equation over energy,
\begin{align}
  \dot\Pi + e^a\D_a\Pi
  &= \underbrace{-4\bigl(\ts\frac13\Theta + A_ae^a
       + \sigma_{ab}e^ae^b\bigr)\Pi}_{\text{redshift}}
  \nonumber\\
  &\quad + \underbrace{K_\perp^a\,\frac{\partial\Pi}{\partial e^a}}_{\text{aberration}}
  \;+\; C[\Pi] .
  \label{eq:liouville}
\end{align}
The redshift term is the integrand of the exact covariant redshift formula of
Zibin and Scott \cite{ZibinScott2008}. The aberration term is its direction
counterpart. Decomposed into PSTF multipoles, Eq.~(\ref{eq:liouville}) splits
every kinematic coefficient of Eq.~(\ref{eq:hierarchy-exact}) into a redshift
part and an aberration part (Table~\ref{tab:coefficients}). The redshift parts
are $O(1)$. Since $\partial O^{A_\ell}/\partial e^a\propto\ell$, the
aberration parts carry the growth:
\begin{equation}
  \bigl(\delta\dot\tau\bigr)_{\!_{NL}}
  = \Bigl[K^a_\perp\,\frac{\partial\tau}{\partial e^a}\Bigr]_{O(\ell)} .
  \label{eq:aberration}
\end{equation}
In PSTF components, moving the $O(\ell)$ terms of
Eq.~(\ref{eq:hierarchy-exact}) to the right-hand side,
\begin{widetext}
\begin{equation}
  \bigl(\delta\dot\tau\bigr)_{\!_{NL}}^{A_\ell} \simeq
  + \ell\Bigl[
    \ts\frac14 \sigma_{bc}\,\tau^{bcA_\ell}
    - \sigma^{\la a_\ell a_{\ell-1}}\tau^{A_{\ell-2}\ra}
    - A^{\la a_\ell}\tau^{A_{\ell-1}\ra}
    + \ts\frac12 A_b\,\tau^{bA_\ell}
    + \omega^b \ep_{bc}{}^{\la a_\ell}\tau^{A_{\ell-1}\ra c}
  \Bigr] .
  \label{eq:nl-source}
\end{equation}
\end{widetext}
The two shear terms carry opposite signs, as the two acceleration terms do. Each
pair is the multipole image of one first-order differential operator on the
sphere of directions, and the top-rank couplings $A\to\ell-1$ and
$\sigma\to\ell-2$ come from the same term of it. Equation~(\ref{eq:nl-source})
is the rate at which the kinematics of the chosen congruence rotate the
direction label relative to the photon. It is not a gravitational interaction
between modes. The term $5\ell/(2\ell+3)\,\sigma_b{}^{\la a_\ell}\Pi^{A_{\ell-1}\ra b}$
has a redshift part and an aberration part that are each $O(1)$ and of
opposite sign.

\subsubsection{The reduction}
\label{sec:reduction}

Two properties of the basis do the whole of the reduction. The first is that a
PSTF coefficient may be contracted on the plain product instead of on the
basis tensor,
\begin{equation}
  \Pi_{A_\ell}O^{A_\ell} = \Pi_{A_\ell}e^{A_\ell} ,
  \label{eq:plain}
\end{equation}
because $O^{A_\ell}=e^{\la A_\ell\ra}$ differs from $e^{A_\ell}$ only by terms
carrying a factor $h^{a_ia_j}$, and $\Pi_{A_\ell}$ is trace-free
[Table~\ref{tab:conventions}, C5.1, C5.8]. The second is the raising form of
the contraction identity,
\begin{equation}
  e^aO^{A_\ell} = O^{aA_\ell}
    + \frac{\ell}{2\ell+1}\,h^{a\la a_\ell}O^{A_{\ell-1}\ra} ,
  \label{eq:eO}
\end{equation}
whose coefficient is fixed by contracting with $e_a$ and using
$\frac{\ell+1}{2\ell+1}+\frac{\ell}{2\ell+1}=1$ [C6.1, C6.3].

Equation~(\ref{eq:plain}) makes the brightness a sum of homogeneous
polynomials in $e^a$, so the direction derivative is the product rule and
nothing else:
\begin{equation}
  \frac{\partial\Pi}{\partial e^a}
  = \sum_{n} n\,\Pi_{aA_{n-1}}e^{A_{n-1}}
  = \sum_{n} n\,\Pi_{aA_{n-1}}O^{A_{n-1}} .
  \label{eq:dPide}
\end{equation}
This is the origin of the $\ell$-growth, and it is one line: differentiating
the direction label brings down a factor equal to the rank, because for this basis
rank \emph{is} polynomial degree. Nothing in it uses a mode expansion, a choice
of harmonic, or proximity to FLRW.

Contracting with $K_\perp^a=K^a-e^a(K_be^b)$ and using
Eq.~(\ref{eq:plain}) backwards on the longitudinal part,
$\Pi_{bA_{n-1}}e^be^{A_{n-1}}=\Pi_{A_n}O^{A_n}$,
\begin{equation}
  K_\perp^a\frac{\partial\Pi}{\partial e^a}
  = \sum_n n\,K^b\Pi_{bA_{n-1}}O^{A_{n-1}}
  - \sum_n n\,(K_be^b)\,\Pi_{A_n}O^{A_n} .
  \label{eq:abersplit}
\end{equation}
The second sum has the same form as the redshift term of
Eq.~(\ref{eq:liouville}) with the weight $-4$ replaced by $-n$. The two
operators differ only in where the weight comes from: the energy integration
supplies a fixed $4$, differentiation supplies the rank. That is the sense in
which Table~\ref{tab:coefficients} records one operator seen twice rather than
a list of coefficients, and it is why no $O(\ell)$ term can appear in the
redshift column.

\subsubsection{The four reductions}
\label{sec:four-reductions}

Putting $K^b$ in turn equal to $A^b$, $\sigma^b{}_ce^c$ and
$\ep^b{}_{cd}e^c\omega^d$ and reducing with Eq.~(\ref{eq:eO}) gives, for the
longitudinal and redshift structures,
\begin{align}
  (A_be^b)\,\Pi_{A_n}O^{A_n}
  &= A^{\la a_{n+1}}\Pi^{A_n\ra}O_{A_{n+1}}
  \nonumber\\
  &\quad + \frac{n}{2n+1}\,A_b\Pi^{bA_{n-1}}O_{A_{n-1}} ,
  \label{eq:red-A}
\end{align}
\begin{align}
  (\sigma_{bc}e^be^c)\,\Pi_{A_n}O^{A_n}
  &= \sigma^{\la a_{n+2}a_{n+1}}\Pi^{A_n\ra}O_{A_{n+2}}
  \nonumber\\
  &+ \frac{2n}{2n+3}\,\sigma_b{}^{\la a_n}\Pi^{A_{n-1}\ra b}O_{A_n}
  \nonumber\\
  &+ \frac{n(n-1)}{(2n+1)(2n-1)}\,\sigma_{bc}\Pi^{bcA_{n-2}}O_{A_{n-2}} ,
  \label{eq:red-s}
\end{align}
and, for the transverse structures,
\begin{align}
  \sigma^b{}_ce^c\,\Pi_{bA_{n-1}}O^{A_{n-1}}
  &= \sigma_b{}^{\la a_n}\Pi^{A_{n-1}\ra b}O_{A_n}
  \nonumber\\
  &\quad + \frac{n-1}{2n-1}\,\sigma_{bc}\Pi^{bcA_{n-2}}O_{A_{n-2}} ,
  \label{eq:red-sT}
\end{align}
\begin{equation}
  \ep^b{}_{cd}e^c\omega^d\,\Pi_{bA_{n-1}}O^{A_{n-1}}
  = \omega^d\ep_{db}{}^{\la a_n}\Pi^{A_{n-1}\ra b}O_{A_n} ,
  \label{eq:red-w}
\end{equation}
while $A^b\Pi_{bA_{n-1}}O^{A_{n-1}}$ is already of rank $n-1$ and needs no
reduction. Equations~(\ref{eq:red-A}), (\ref{eq:red-sT}) and (\ref{eq:red-w})
are one application of Eq.~(\ref{eq:eO}), the coefficient tensor being
trace-free on the indices the bracket acts on, so the bracket may be dropped
against it. Equation~(\ref{eq:red-w}) has no rank-lowering term because that
term would carry $\Pi^{bc}$ contracted on the antisymmetric pair of
$\ep_{bcd}$: vorticity couples $\ell$ to itself and to nothing else, and this
is the reason.

Equation~(\ref{eq:red-s}) needs Eq.~(\ref{eq:eO}) twice and one trace removal.
Applying it once with the free index left on $\Pi$,
\begin{equation}
  \Pi_{A_n}e^dO^{A_n} = \Pi_{A_n}O^{dA_n}
    + \frac{n}{2n+1}\,\Pi^d{}_{A_{n-1}}O^{A_{n-1}} ,
  \label{eq:eO-Pi}
\end{equation}
and contracting with $\sigma_{cd}e^c$. The second term reduces by
Eq.~(\ref{eq:eO}) again --- its coefficient $\sigma_{cd}\Pi^d{}_{A_{n-1}}$ is
trace-free on $A_{n-1}$ --- and contributes $\frac{n}{2n+1}$ to the
rank-$n$ structure and $\frac{n}{2n+1}\frac{n-1}{2n-1}$ to the rank-$(n-2)$
one, which is the third coefficient of Eq.~(\ref{eq:red-s}). The first term
needs the trace-free part of $h^{c\la d}O^{A_n\ra}$. For a symmetric rank-$m$
tensor $T$ whose trace $t^{B_{m-2}}=T^{B_{m-2}c}{}_c$ is itself trace-free,
\begin{equation}
  T^{\la B_m\ra} = T^{B_m}
    - \frac{m(m-1)}{2(2m-1)}\,h^{(b_1b_2}t^{B_{m-2})} ,
  \label{eq:detrace}
\end{equation}
the coefficient following from
$h_{b_1b_2}h^{(b_1b_2}t^{B_{m-2})}=\frac{2(2m-1)}{m(m-1)}t^{B_{m-2}}$, in which
$2m-1$ of the $\binom{m}{2}$ index pairings of the symmetrisation survive the
trace and the rest die on the trace-freeness of $t$ (App.~\ref{app:longer}).
With $m=n+1$ and $T^{dA_n}=v^{(d}O^{A_n)}$, whose trace is
$\frac{2}{n+1}v_bO^{bA_{n-1}}$, Eq.~(\ref{eq:detrace}) gives
\begin{equation}
  v^{\la d}O^{A_n\ra} = v^{(d}O^{A_n)}
    - \frac{n}{2n+1}\,h^{(da_1}v_bO^{A_{n-1})b} .
  \label{eq:vO}
\end{equation}
Setting $v^d=h^{cd}$ and contracting with $\sigma_{cd}\Pi_{A_n}$, the pairing
that puts $h^{cd}$ on $\sigma$ dies on $\sigma^c{}_c=0$ and every pairing
internal to $A_n$ dies on the trace-freeness of $\Pi$, leaving $n$ surviving
pairings in the first term and $n$ of the $\binom{n+1}{2}$ in the second, so
that the rank-$n$ part of $\sigma_{cd}\Pi_{A_n}e^cO^{dA_n}$ is
\begin{equation}
  \frac{n+1}{2n+3}\Bigl[\frac{n}{n+1}-\frac{2n}{(2n+1)(n+1)}\Bigr]
  = \frac{n(2n-1)}{(2n+3)(2n+1)} .
\end{equation}
Adding the $\frac{n}{2n+1}$ from the other branch,
\begin{equation}
  \frac{n}{2n+1} + \frac{n(2n-1)}{(2n+3)(2n+1)} = \frac{2n}{2n+3} ,
\end{equation}
the coefficient printed in Eq.~(\ref{eq:red-s}).

\subsubsection{Collecting}
\label{sec:collecting}

Every entry of Table~\ref{tab:coefficients} now follows. A coupling at rank
$\ell$ draws on exactly one source rank $n$: the rank of the multipole it
couples in. Writing $r$ for the coefficient with which that structure appears
in Eq.~(\ref{eq:red-A}) or~(\ref{eq:red-s}), and $t$ for minus the coefficient
with which it appears in the transverse sum of Eq.~(\ref{eq:abersplit}), the
redshift entry is $4r$ and the aberration entry is $nr+t$, the signs being
those of the left-hand side of Eq.~(\ref{eq:hierarchy-exact}). The vorticity
row has $r=0$ because $\ep_{bcd}e^be^c\omega^d$ vanishes identically, which is
why it has no redshift part at all.

Two readings of the table are worth stating. The top-rank couplings
$A\to\ell-1$ and $\sigma\to\ell-2$ are the rows with $r=1$ and $t=0$, so each
is $4+n$ with $n$ its own source rank; they carry the same sign because the
leading term of a reduction has coefficient one whatever the rank, not because
of any cancellation. And the $O(\ell)$ growth sits entirely in $nr+t$: $4r$ is
bounded for every row, so no choice of threading can move the growth into the
redshift column.

\begin{table*}[tp]
\caption{Each kinematic coefficient of Eq.~(\ref{eq:hierarchy-exact})
(left-hand side), derived. A coupling at rank $\ell$ draws on one source rank
$n$, the rank of the multipole it couples in, and appears in the reduction of
Eqs.~(\ref{eq:red-A})--(\ref{eq:red-s}) with coefficient $r$ and in the
transverse sum of Eq.~(\ref{eq:abersplit}) with coefficient $-t$. The redshift
entry is then $4r$, the aberration entry $nr+t$, and their sum is the
coefficient printed in Eq.~(\ref{eq:hierarchy-exact}). Here
$c_2\equiv(\ell+1)(\ell+2)/[(2\ell+3)(2\ell+5)]$. Every $O(\ell)$ coefficient
is aberration: $4r$ is bounded in every row. The vorticity row has $r=0$
because $\ep_{bcd}e^be^c\omega^d\equiv0$, and the orientation of $\ep_{abc}$ is
a stipulation, not a result. Notes in Sec.~\ref{sec:collecting}.}
\label{tab:coefficients}
\begin{ruledtabular}
\begin{tabular}{lcccccccc}
Term & $n$ & $r$ & $t$ & Redshift & Aberration & Coefficient & $\ell\gg1$ \\
 & & & & $4r$ & $nr+t$ & $4r+nr+t$ & \\
\colrule
$A^{\la a_\ell}\Pi^{A_{\ell-1}\ra}$
  & $\ell-1$ & $1$ & $0$
  & $+4$ & $+(\ell-1)$ & $+(\ell+3)$ & $+\ell$ \\[3pt]
$A_b\Pi^{bA_\ell}$
  & $\ell+1$ & $\dfrac{\ell+1}{2\ell+3}$ & $-(\ell+1)$
  & $+\dfrac{4(\ell+1)}{2\ell+3}$ & $-\dfrac{(\ell+1)(\ell+2)}{2\ell+3}$
  & $-\dfrac{(\ell+1)(\ell-2)}{2\ell+3}$ & $-\ell/2$ \\[8pt]
$\sigma^{\la a_\ell a_{\ell-1}}\Pi^{A_{\ell-2}\ra}$
  & $\ell-2$ & $1$ & $0$
  & $+4$ & $+(\ell-2)$ & $+(\ell+2)$ & $+\ell$ \\[3pt]
$\sigma_{bc}\Pi^{bcA_\ell}$
  & $\ell+2$ & $c_2$ & $-\dfrac{(\ell+1)(\ell+2)}{2\ell+3}$
  & $+4c_2$ & $-(\ell+3)c_2$ & $-(\ell-1)c_2$ & $-\ell/4$ \\[8pt]
$\sigma_b{}^{\la a_\ell}\Pi^{A_{\ell-1}\ra b}$
  & $\ell$ & $\dfrac{2\ell}{2\ell+3}$ & $-\ell$
  & $+\dfrac{8\ell}{2\ell+3}$ & $-\dfrac{3\ell}{2\ell+3}$
  & $+\dfrac{5\ell}{2\ell+3}$ & $O(1)$ \\[8pt]
$\omega^b\ep_{bc}{}^{\la a_\ell}\Pi^{A_{\ell-1}\ra c}$
  & $\ell$ & $0$ & $-\ell$
  & $0$ & $-\ell$ & $-\ell$ & $-\ell$ \\
\end{tabular}
\end{ruledtabular}
\end{table*}

Both rates of Eqs.~(\ref{eq:dE})--(\ref{eq:de}) come from
$p^b\nabla_bu^a = E(A^a + e^b\nabla_bu^a)$. The $\frac13\Theta e^a$ part of
$e^b\nabla_bu^a$ lies along $e^a$ and does not rotate the direction, which is
why $\Theta$ appears in the redshift column of Eq.~(\ref{eq:liouville}) and
nowhere in $K^a$. The Liouville equation \cite{EllisMatraversTreciokas1983,
MaartensGebbieEllis1999}, integrated by parts in $E$, gives
Eq.~(\ref{eq:liouville}).

The five $A$ and $\sigma$ rows of the table are derived above from the
Liouville operator alone, with no input from the hierarchy they are being
compared against. All five agree with Eq.~(70) of
Ref.~\cite{MaartensGebbieEllis1999} integrated over energy, and all but the
$\ell-2$ row with its Eq.~(71). The
$\ell-2$ row agrees with the low-$\ell$ Eqs.~(87) and (89) of the same paper,
which fix the $A$ and $\ell-2$ shear signs at $\ell=1,2$. For the
rank-preserving shear term at $\ell=2$, Eq.~(89) prints $\frac27$, where
Eq.~(70) integrated over energy and the decomposition here give
$\frac{10}7$; the term is $O(1)$ and is not used. The vorticity row is taken
from Eq.~(70); its sign depends on the orientation of $\ep_{abc}$.

The top-rank couplings $A\to\ell-1$ and $\sigma\to\ell-2$ are the two rows of
Table~\ref{tab:coefficients} with $r=1$ and $t=0$: each comes from the leading
term of one reduction and from one term of the redshift, and each is therefore
$4+n$. The $\ell-2$ shear coupling rests on that identity and on nothing else,
which is the subject of the next section.

\subsection{The top-rank couplings and the sign of the source}
\label{sec:shearsign}

The last coupling of Eq.~(\ref{eq:hierarchy-exact}),
$+(\ell+2)\sigma^{\la a_\ell a_{\ell-1}}\Pi^{A_{\ell-2}\ra}$, carries the same
plus as the $\ell-1$ acceleration coupling for a reason belonging to neither
term separately: both are the leading term of a reduction, and a leading term
has coefficient one whatever its rank. Section~\ref{sec:liouville} obtains it at
general $\ell$ from $p^b\nabla_bp^a=0$, through the Liouville operator and the
energy integration, taking nothing from the printed hierarchy and reproducing
every other coefficient in it.

What the sign settles is which kind of operator the coupling is. The two shear
terms carry opposite signs, as the two acceleration terms do. A common sign
would make the shear coupling a growth rate proportional to $\ell$ in every
multipole. Opposite signs make it a remapping of directions, which moves power
between neighbouring multipoles (Sec.~\ref{sec:aligned}). The rank-preserving
coupling $\sigma_b{}^{\la a_\ell}\Pi^{A_{\ell-1}\ra b}$ tends to $\frac52$ and
is not part of the $O(\ell)$ source.

\subsubsection{Transposition}
\label{sec:transposition}

Moving the $O(\ell)$ terms of Eq.~(\ref{eq:hierarchy-exact}) to the right-hand
side gives Eq.~(\ref{eq:nl-source}). Two facts fix its signs without appeal to
any later equation. First, every coefficient in Table~\ref{tab:coefficients} is
a left-hand-side coefficient, and transposing a term of coefficient $c\,\ell$
gives a source $-c\,\ell$. The high-$\ell$ limits $+\ell$, $-\ell/2$, $+\ell$,
$-\ell/4$ and $-\ell$ of the $A\to\ell-1$, $A\to\ell+1$, $\sigma\to\ell-2$,
$\sigma\to\ell+2$ and $\omega$ couplings therefore give the coefficients $-1$,
$+\frac12$, $-1$, $+\frac14$ and $+1$ of Eq.~(\ref{eq:nl-source}). Second,
passing from $\Pi^{A_\ell}$ to $\tau^{A_\ell}$ divides by a positive
normalisation, $\Pi^{A_\ell}\propto\rho_r\tau^{A_\ell}$, and cannot change the
sign of a term linear in $\Pi^{A_\ell}$. The overall sign of the bracket is
$+\ell$.

One further route to $+(\ell+2)$ at general $\ell$, and two that agree at
$\ell=2$, are set out in the supplementary material \cite{Bundle}.

\section{The weakly nonlinear system at $O(\ep^2\ell)$}
\label{sec:weaklynl}

Linearising the hierarchy about FLRW removes every product of two first-order
quantities, and the kinematic couplings are all of that form. This section
isolates what is removed, writes it in an arbitrary threading, and reduces the
four couplings that survive to PSTF form. Nothing here is an approximation to
the exact hierarchy of Sec.~\ref{sec:hierarchy}; it is a sorting of its terms by
order in the perturbation amplitude.

\subsection{What linearisation removes, and why it has to be put back}
\label{sec:linearisation-removes}

Every kinematic coupling in Eq.~(\ref{eq:hierarchy-exact}) is a product of a
kinematic quantity with an anisotropy multipole, and both factors vanish in the
exact Friedmann--Lema\^itre limit. Each is therefore $O(\ep^2)$
[Sec.~\ref{sec:orders}] and each is removed by linearising about FLRW. The two
terms in $v^a_{\!_B}$ on the right of Eq.~(\ref{eq:hierarchy-exact}) are
quadratic in the same sense and go with them; they are the
$\delta\dot C_{\!_{NL}}$ of Eq.~(\ref{eq:master}) below. What is left is the
free-streaming operator, the expansion damping $\frac43\Theta\Pi^{A_\ell}$ ---
in which $\Theta$ multiplies the anisotropy but is itself a background quantity
--- and the linear Thomson term.

Collected, the removed terms are the aberration term of the Liouville equation,
$K^a_\perp\,\partial\Pi/\partial e^a$, with $K^a$ the single contraction of
Eq.~(\ref{eq:K}); Table~\ref{tab:coefficients} is that one term written out in
multipoles. What linearisation drops is a single operator and not five
independent couplings: the rate at
which the threading's direction label turns relative to the photon.

The same kinematic quantities survive linearisation elsewhere. In
Eq.~(\ref{eq:B}) the acceleration and the shear
appear again, contracted with $e^a$ alone; there they multiply the monopole
rather than an anisotropy, so the product is first order and is kept. That term
is the covariant integrated redshift --- Sachs--Wolfe and Rees--Sciama. It is
not the coupling discussed here, and the two are easily conflated because the
same $\sigma_{ab}$ stands in both.

The removed operator is the only term of Eq.~(\ref{eq:hierarchy-exact}) that
differentiates the direction label, and that is why it grows with $\ell$: by
Eq.~(\ref{eq:dPide}) such a derivative brings down a factor equal to the rank, because
for this basis rank is polynomial degree. Every coefficient that survives
linearisation is bounded as $\ell\to\infty$ --- $1$ and $(\ell+1)/(2\ell+3)$ for
free streaming, $\frac43\Theta$ for the damping --- and every coefficient that
grows with $\ell$ is removed. The trade is therefore not a fixed number. What is
dropped is smaller than what is kept by one power of the amplitude and larger by
one power of the rank, and $\ep\ell$ at $\ell\sim10^3$ is not small in the way
it is at $\ell\sim10$. The rank-preserving $5\ell/(2\ell+3)$ is the one removed
coupling that is $O(1)$, and it is also the one that does not survive the
high-$\ell$ limit of Eq.~(\ref{eq:nl-source}).

It has to be put back because it is lensing. $K^a_\perp$ measures the deflection
of the ray by the intervening field together with the aberration of the
observer's own sky, and Sec.~\ref{sec:content} identifies its $O(\ep^2\ell)$
content as those two and nothing else. Removing it from the hierarchy removes
them from the solution, and they are restored afterwards by another route: solve
the linearised hierarchy, then remap the resulting sky with a deflection taken
from the geodesic equation \cite{LewisChallinor2006}. That is the right way to
compute, for the reason given in Sec.~\ref{sec:discussion}, and nothing here
argues otherwise. What the removal takes away is a statement, not a number. A quantity the
hierarchy already contains has to be re-imported from \emph{outside} it; and a
threading is fixed in the course of writing the linear equations down, after
which the frame covariance of Sec.~\ref{sec:frame} --- which is the reason the
two reappear as one object --- is no longer displayed within the reduced system.

That last point is not a feature of the practice but of the setting it is carried
out in. Defining a perturbation at all requires a background and a map
identifying points of the spacetime with points of it, and fixing that map
selects a foliation: what a change of physical threading would redistribute is
then not visible as an internal covariance of the reduced equations, and has to
be recovered by an explicit observer or tetrad transformation. The theory is not
less covariant for that; the reduced description is, and that is why lensing is
brought back into it from outside. The $1+3$ variables need no such map, and
that is the point of working in them.

\subsection{The $O(\ep^2\ell)$ source in an arbitrary threading}
\label{sec:arbitrary-threading}

Collecting terms by order, the temperature anisotropy obeys
\begin{equation}
  \dot\tau + e^a\D_a\tau \simeq
  \underbrace{\mathcal{B} + C[\tau]}_{\text{almost-FLRW}}
  + \underbrace{(\delta\dot\tau)_{\!_{OV}}
    + (\delta\dot\tau)_{\!_{NL}}
    + \delta\dot C_{\!_{NL}}}_{O(\ep^2)} ,
  \label{eq:master}
\end{equation}
with the linear redshift source and Thomson term
\begin{align}
  \mathcal{B}(x,e) &\simeq +\ts\frac13\D^a\tau_a
    - (\D_a\ln T + A_a)e^a - \sigma_{ab}e^ae^b ,
  \label{eq:B}\\
  C[\tau](x,e) &\simeq +\st\ne\bigl(\bar f_1 e^a v^B_a
    + \bar f_2 e^ae^b\tau_{ab} - \tau\bigr) ,
\end{align}
the Ostriker--Vishniac source
$(\delta\dot\tau)_{\!_{OV}} \simeq +\st\ne\,\delta_{\!_E}\bar f_1e^av^B_a$, the aberration
source $(\delta\dot\tau)_{\!_{NL}}$ of Eq.~(\ref{eq:nl-source}), and the
nonlinear Thomson correction $\delta\dot C_{\!_{NL}}$. The free-streaming
operator $e^a\D_a$ on the left acts on the threading's perturbed spatial
sections, and it is not linearised. That matters in Sec.~\ref{sec:lensing}.

In the free-streaming era the vorticity is negligible for scalar modes. In the
CDM threading, $A_a\approx0$, the source reduces to
\begin{equation}
  \bigl(\delta\dot\tau\bigr)^{A_\ell}_{\!_{NL}} \simeq
  +\,\ell\Bigl[\ts\frac14\sigma_{bc}\tau^{bcA_\ell}
  - \sigma^{\la a_\ell a_{\ell-1}}\tau^{A_{\ell-2}\ra}\Bigr] ,
  \label{eq:source-cdm}
\end{equation}
and in the Newtonian threading, $\tilde\sigma_{ab}=0$, to the two acceleration
terms. The nonlinear Thomson correction of Eq.~(\ref{eq:master}) is
\begin{equation}
  \delta\dot C_{\!_{NL}} \simeq +\st\ne\,O_{A_\ell}
    \bigl[\tau^{\la A_{\ell-1}}v^{a_\ell\ra}_B
    + \ts\frac12\tau^{A_\ell a}v^B_a\bigr] .
\end{equation}
Neither it nor the Ostriker--Vishniac term is pursued here; the latter is the
perturbation of $\ne$ in the Thomson term of Eq.~(\ref{eq:master}), with
$\delta_{\!_E}$ the electron density contrast. The shear also
enters the linear source $\mathcal B$ through $\sigma_{ab}e^ae^b$, the
covariant integrated redshift; that is a different term from the coupling of
Eq.~(\ref{eq:source-cdm}).

\subsection{PSTF reduction of the four couplings}
\label{sec:pstfreduction}

The four weights below are contraction identities and nothing else. They are
read in the \emph{aligned} configuration --- every PSTF tensor built on one
common unit vector $\hat k$, as Sec.~\ref{sec:aligned} specifies covariantly ---
so each weight is one of the identities of Sec.~\ref{sec:conv-pstf} with
$\hat k$ in place of $e$: the two rank-lowering weights are C6.1 and C6.2,
\begin{equation}
\begin{gathered}
  \hat k^a\hat k_{\la aA_\ell\ra} = \frac{\ell+1}{2\ell+1}\,\hat k_{\la A_\ell\ra} ,
  \\
  \hat k^a\hat k^b\hat k_{\la abA_\ell\ra}
   = \frac{(\ell+1)(\ell+2)}{(2\ell+1)(2\ell+3)}\,\hat k_{\la A_\ell\ra} ,
\end{gathered}
  \label{eq:kcontr}
\end{equation}
and the two rank-raising weights are unity because the leading term of the PSTF
recursion is, as in Eq.~(\ref{eq:red-A}), of coefficient one. No
specialisation to $\ell=2$ is used and nothing is settled by a mode expansion.

\emph{What the aligned reading does and does not supply.} Equation~(\ref{eq:kcontr})
is an identity between tensors built on a common $\hat k$. It is not the
contraction of the harmonics of Eq.~(\ref{eq:harmonic}): those carry a
prefactor that depends on rank through the derivatives of the eigenfunction, so
a contraction that lowers the rank by two changes its sign relative to the
algebraic weight, and a product of two modes is in general not aligned with the
resultant wavevector. The spatial convolution kernel therefore does not follow
from Eq.~(\ref{eq:kcontr}) and is not derived here. What follows below is the
aligned-amplitude form, written with the mode labels for readability, and it is
used only for the order-of-magnitude weighting of
Sec.~\ref{sec:scalefactor} and for the comparison in
Sec.~\ref{sec:aligned}; nothing in Secs.~\ref{sec:content}
or~\ref{sec:frame} rests on it.

With that reading, the four distinct couplings carry
\begin{widetext}
\begin{align}
  A_{\la a_\ell}\tau_{A_{\ell-1}\ra}
   &\sim (4\pi)\,Q_{A_\ell}\!\int\!\frac{k^2\dd k}{(2\pi)^3}\,
     A(k)\,\tau_{\ell-1}(|k^*-k|) ,
   &&\text{coefficient } 1 ,
   \label{eq:c1}\\
  A^a\tau_{aA_\ell}
   &\sim (4\pi)\,\frac{(\ell+1)}{(2\ell+1)}\,Q_{A_\ell}\!\int\!\frac{k^2\dd k}{(2\pi)^3}\,
     A(k)\,\tau_{\ell+1}(|k^*-k|) ,
   &&\to \tfrac12 ,
   \\
  \sigma_{\la a_\ell a_{\ell-1}}\tau_{A_{\ell-2}\ra}
   &\sim (4\pi)\,Q_{A_\ell}\!\int\!\frac{k^2\dd k}{(2\pi)^3}\,
     \sigma(k)\,\tau_{\ell-2}(|k^*-k|) ,
   &&\text{coefficient } 1 ,
   \\
  \sigma^{ab}\tau_{abA_\ell}
   &\sim (4\pi)\,\frac{(\ell+1)(\ell+2)}{(2\ell+1)(2\ell+3)}\,Q_{A_\ell}\!
     \int\!\frac{k^2\dd k}{(2\pi)^3}\,
     \sigma(k)\,\tau_{\ell+2}(|k^*-k|) ,
   &&\to \tfrac14 .
   \label{eq:c4}
\end{align}
\end{widetext}
Combined with the hierarchy limits of Table~\ref{tab:coefficients}, the
coefficient on $\tau_{\ell+2}$ is $\frac14\times\frac14=\frac1{16}$ and that on
$\tau_{\ell+1}$ is $\frac12\times\frac12=\frac14$: in each case one factor from
the hierarchy and one from the contraction. That factorisation is exact and not
only a limit, and it can be checked without this reduction at all. Multiplying
the aberration entries of Table~\ref{tab:coefficients} by the weights of
Eq.~(\ref{eq:kcontr}),
\begin{equation}
\begin{gathered}
  \frac{(\ell+1)(\ell+2)}{2\ell+3}\cdot\frac{\ell+1}{2\ell+1}
   = \frac{(\ell+1)^2(\ell+2)}{(2\ell+1)(2\ell+3)} ,
  \\
  (\ell+3)c_2\cdot\frac{(\ell+1)(\ell+2)}{(2\ell+1)(2\ell+3)}
   = \frac{(\ell+1)^2(\ell+2)^2(\ell+3)}{(2\ell+1)(2\ell+3)^2(2\ell+5)} ,
\end{gathered}
  \label{eq:factorise}
\end{equation}
and the right-hand sides are Eqs.~(\ref{eq:alignN}) and~(\ref{eq:alignC}),
which Sec.~\ref{sec:aligned} obtains in index algebra without passing through a
mode expansion. The rank-lowering couplings factorise the same way with weight
one, giving $-(\ell-1)$ and $-(\ell-2)$. The two routes agree term by term and
at every $\ell$, which is what closes this block. The source in mode form is
\begin{equation}
\begin{split}
  \frac{\dd}{\dd t}&(\delta\tau)^{\!_{NL}}_\ell(k^*,t)
  \simeq +\,(4\pi)\,\ell\!\int_{k_0}^{\infty}\!\frac{k^2\dd k}{(2\pi)^3} \\
  &\times\Bigl[-\sigma(k,t)\,\tau_{\ell-2}(|k^*-k|,t) \\
  &\qquad\ + \ts\frac{1}{16}\sigma(k,t)\,\tau_{\ell+2}(|k^*-k|,t) \\
  &\qquad\ - A(k,t)\,\tau_{\ell-1}(|k^*-k|,t) \\
  &\qquad\ + \ts\frac14 A(k,t)\,\tau_{\ell+1}(|k^*-k|,t)\Bigr] .
\end{split}
  \label{eq:modeform}
\end{equation}

\subsection{The scale factor}
\label{sec:scalefactor}

The source (\ref{eq:modeform}) is a rate per unit proper time, since $\sigma$
and $A$ are proper-time rates, so it is integrated against
$\dd t = a\,\dd\eta$. In the CDM threading,
\begin{align}
  (\delta\tau)^{\!_{NL}}_\ell(k^*,\eta_0) &\simeq
  +\,4\pi\ell\!\int_{\eta_*}^{\eta_0}\!\!a(\eta)\,\dd\eta
  \int_{k_0}^{\infty}\!\frac{k^2\dd k}{(2\pi)^3}\,\sigma(k,\eta)
  \nonumber\\
  &\quad\times\Bigl[-\mathcal{C}_-(\ell)\,\tau_{\ell-2}(|k^*-k|,\eta)
  \nonumber\\
  &\qquad\ \ + \mathcal{C}_+(\ell)\,\tau_{\ell+2}(|k^*-k|,\eta)\Bigr] ,
  \label{eq:nlrs}
\end{align}
with $\mathcal{C}_\mp>0$, equal to $1$ and $\frac1{16}$ in the aligned limit.
The shear is fixed by the potentials. The \emph{energy frame} is the threading
in which the total energy flux vanishes, $q_a=0$; it and the CDM threading of
Sec.~\ref{sec:threading} are separate stipulations, and they agree only to
$\approx$ [Table~\ref{tab:conventions}, C0.1]. Setting $q_a=0$ in a mixture
gives $\rho_{\!_C}v^a_{\!_C}+\frac43\rho_{\!_R}v^a_{\!_R}+\ldots=0$, so the
energy frame tilts away from the CDM one by $O(\rho_{\!_R}/\rho_{\!_C})$ and the
two coincide in the limit of matter domination; the smallness parameter is the
ratio of radiation to matter density, which is the one
Ref.~\cite{GebbieDunsbyEllis1999} names for $\approx$. A cosmological constant
does not enter the comparison at all: $\rho_\Lambda+p_\Lambda=0$ and
$q^\Lambda_a=0$, so $\Lambda$ contributes neither to the flux that defines the
energy frame nor to the inertia that weights it. The identification is a
statement about the radiation fraction, not about whether the model carries a
cosmological constant. With that understood, Eq.~(29) of
Ref.~\cite{GebbieDunsbyEllis1999} gives, its flux term dropped by the frame
choice,
\begin{equation}
\begin{split}
  \ts\frac12(\rho+p)\,\sigma_{ab} \simeq{}&
  \bigl(\D_{\la a}\D_{b\ra}\Phi_H\bigr)^{\!\textbf{.}}
  + 3H\,\D_{\la a}\D_{b\ra}\Phi_H \\
  &- H\,\D_{\la a}\D_{b\ra}(\Phi_H+\Phi_A) ,
\end{split}
\end{equation}
and in matter domination ($A_a\approx0$, $\Phi_H\approx-\Phi_A$), with
$\D_a\Phi_H = +k\,\Phi_H Q_a$ and
$\D_{\la a}\D_{b\ra}\Phi_H = -k^2\Phi_H Q_{ab}$ with $k$ physical,
\begin{equation}
  \sigma(k,\eta) \simeq -\frac{D'}{a}\,\Delta(k,\eta_0) ,
  \label{eq:shear-source}
\end{equation}
consistent with $\frac32H_0^2\Omega_0\Delta = -k^2\Phi_A(k,0) = +k^2\Phi_H(k,0)$,
evaluated today, where the physical wavenumber and its comoving label coincide.
The Jacobian cancels the $1/a$ in the shear exactly,
\begin{equation}
  a(\eta)\,\sigma(k,\eta)\,\dd\eta \simeq -\Delta(k,\eta_0)\,\dd D ,
  \label{eq:aweight}
\end{equation}
so the kinematic weight of the shear along the line of sight is uniform in the
growth factor $D$. In Einstein--de~Sitter, $D=a=(\eta/\eta_0)^2$, the integral
is $-\Delta(k,\eta_0)(1-a_*)$, and half of it accrues at $z<1$.

\section{The $O(\ep^2\ell)$ content is lensing and aberration}
\label{sec:content}
\label{sec:physics}
What this section establishes is the frame-covariant form. That the
$O(\ep^2\ell)$ content is gravitational lensing, and that the coefficients grow
in $\ell$, are known in metric variables \cite{BenekeFidler2010}, and are cited
to that work; what is shown here is that both halves of the deflection and the
observer's own aberration are carried by one operator whose threading
dependence is the relabelling of Sec.~\ref{sec:relabel}.

With the endpoints fixed, the observed sky depends on three things only: the
temperature and frame of the emitter, the null geodesics connecting emitter to
observer, and the observer's four-velocity $u_R^a$. Expanding the observed
temperature to $O(\ep^2)$ along the true geodesic and listing every product
of two $O(\ep)$ quantities sorts the standard secondary anisotropies by
order, and by the term of Eq.~(\ref{eq:master}) that carries each
(Table~\ref{tab:physics}). The organising distinction is whether a transverse
or angular gradient acts on the high-$\ell$ pattern of $\tau$. If it does, the
coupling is $O(\ep^2\ell)$, with $\ep$ the perturbation amplitude; if
not, it is $O(\ep^2\ell^0)$.

\begin{table*}[tp]
\caption{The standard CMBR secondaries in the $1+3$ hierarchy, by order and by
the term of Eq.~(\ref{eq:master}) or of the geodesic map that carries each.
$O(\ep^2\ell)$ means a transverse or angular gradient acting on $\tau$.
The enumeration is described in Sec.~\ref{sec:residue}.}
\label{tab:physics}
\begin{ruledtabular}
\begin{tabular}{>{\raggedright\arraybackslash}p{0.17\textwidth}l>{\raggedright\arraybackslash}p{0.34\textwidth}>{\raggedright\arraybackslash}p{0.30\textwidth}}
Effect & Order & Carried by & Comment \\
\colrule
Sachs--Wolfe, Doppler, ISW
  & $\ep$ & redshift source $\mathcal B$, Eq.~(\ref{eq:B}); endpoints
  & Ref.~\cite{ChallinorLasenby1998} \\
Rees--Sciama, moving lens
  & $\ep^2\ell^0$ & redshift term of Eq.~(\ref{eq:liouville}) at $O(\ep^2)$
  & gradient acts on $\Phi$, not on $\tau$ \\
Ostriker--Vishniac
  & $\ep^2$ & $(\delta\dot\tau)_{\!_{OV}}$, Thomson scattering
  & a source, not a coupling to $\tau_\ell$ \\
Time delay
  & $\ep^2\ell^0$ & radial displacement in the geodesic map; no $\ell$ while
    the dominant modes are nearly transverse
  & Ref.~\cite{HuCooray2001} \\
Aberration at emission
  & $\ep^2\ell^0$ & $K_\perp^a\partial\tau/\partial e^a$ acting on $\ell\le2$ at $E$
  & suppressed by tight coupling \cite{ZibinScott2008} \\
Observer modulation
  & $\ep^2\ell^0$ & Doppler factor at $R$
  & Ref.~\cite{Planck2013XXVII} \\
Observer aberration
  & $\ep^2\ell$ & $v_{R\perp}^a\partial\tau/\partial e^a$ at $R$
  & Ref.~\cite{Planck2013XXVII} \\
Lensing, $\Phi_A$ half
  & $\ep^2\ell$ & $K_\perp^a\partial\tau/\partial e^a$, Eq.~(\ref{eq:aberration})
  & equivalence principle; Sec.~\ref{sec:lensing} \\
Lensing, $\Phi_H$ half
  & $\ep^2\ell$ & $e^a\D_a\tau$ on the perturbed sections, Eq.~(\ref{eq:curvop})
  & spatial curvature; Sec.~\ref{sec:curvhalf} \\
Lensing by tensors
  & $\ep^2\ell$ & tensor shear in $K_\perp^a$
  & same operator and $\ell$-weight; amplitude set by the tensor shear \\
Post-Born, curl deflection
  & $\ep^3$ & deflection of the deflection
  & beyond this order, \emph{for scalar modes}: at Born order a scalar
    deflection is the gradient of $\psi$ and has no curl, while a tensor shear
    gives the deflection a curl part at the order of its gradient part \\
Baryon velocity beyond linear
  & $\ep\vb^2$, $\vb^3$ & Thomson term of Eq.~(\ref{eq:hierarchy-exact})
  & the one truncation there; counted in $\vb$, not in $\ep$ \\
\end{tabular}
\end{ruledtabular}
\end{table*}

\subsection{First order and $O(\ep^2\ell^0)$}
\label{sec:first-order}

At linear order the right-hand side of Eq.~(\ref{eq:master}) is
$\mathcal B + C[\tau]$. The source $\mathcal{B}$ is the linearised redshift term
of Eq.~(\ref{eq:liouville}): the expansion, acceleration and shear of the
threading, projected on the line of sight. Integrated along the ray with the
Thomson term, it yields the Sachs--Wolfe, Doppler and integrated Sachs--Wolfe
(ISW) contributions, in $1+3$ variables, as in Ref.~\cite{ChallinorLasenby1998}. The
aberration term of Eq.~(\ref{eq:liouville}) does not appear: acting on the
isotropic background it vanishes, and acting on the anisotropy it is second
order.

The redshift term of Eq.~(\ref{eq:liouville}) at $O(\ep^2)$ carries the
Rees--Sciama effect, including the moving-lens term
$v_\perp\cdot\nabla_{\!\perp}\Phi_A$. Its gradients act on the potential, not
on the anisotropy, so it carries no factor of $\ell$; the standard scaling
argument \cite{Seljak1996, MartinezGonzalezSanzSilk1994} applies to it as it
stands. The Ostriker--Vishniac effect is a scattering source in
$(\delta\dot\tau)_{\!_{OV}}$. The time delay, the observer's Doppler modulation,
and the aberration of the emitted pattern at last scattering complete the list.
The last acts on a field whose direction dependence at fixed position has
$\ell\le2$ in the baryon frame, so its coupling is $O(1)$ even though it comes
from the aberration operator. The $O(\ep^2\ell^0)$ terms are
frame-covariant in combination by the exact result of
Ref.~\cite{ZibinScott2008}, and they are not pursued here.

\subsection{The aberration operator and the equivalence-principle half}
\label{sec:lensing}
\begin{figure}[!htbp]
\includegraphics[width=\columnwidth]{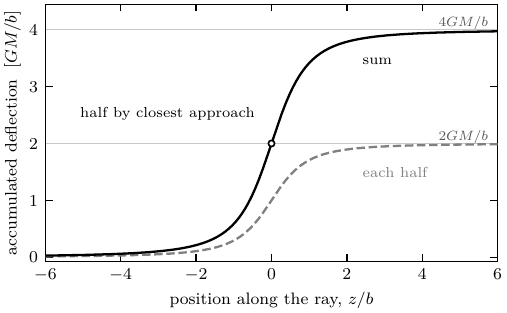}
\caption{The two halves of the deflection, accumulated along the ray past a
point mass at impact parameter $b$. The equivalence-principle half is carried by
the aberration operator [Eq.~(\ref{eq:bendK})] and the spatial-curvature half by
free streaming on the perturbed sections [Eq.~(\ref{eq:bendH})]. They are not
merely equal in total: with the anisotropic stress vanishing,
$\Phi_H=-\Phi_A$ [Table~\ref{tab:conventions}, C2.5], so the two are the same
function of position and one curve carries both. Each tends to $2GM/b$ and their
sum to $4GM/b$, which is the completion the hierarchy must supply. The sum passes
$2GM/b$ exactly at closest approach, so half the bending is laid down on the way
in and half on the way out. It is not an endpoint effect --- $71\%$ of it falls
within $|z|<b$ --- which is what distinguishes it from the endpoint terms of
Sec.~\ref{sec:bulkendpoint}.}
\label{fig:halves}
\end{figure}

Two terms of the geodesic map grow as $\ell$: the deflection of the ray, which
displaces the emission point, and the aberration of the sky by $u^a_R$ relative
to the frame in which $\tau$ is computed. The second is the observer's kinematic
aberration. It has been measured \cite{Planck2013XXVII}, and on the full sky it
couples $\ell$ only to $\ell\pm1$, leaving $C_\ell$ unchanged at first order in
the observer's velocity. The first is gravitational lensing. We derive it in the
Newtonian threading, where $\tilde\sigma_{ab}=0=\tilde\omega_a$ and
$\tilde A_a=\D_a\Phi_A$. Write $\nabla_{\!\perp}\equiv a\D_\perp$ for the
comoving transverse gradient, $\chi$ for comoving distance back from the
observer along the background ray, and $\chi_*$ for the distance to last
scattering.

\emph{The equivalence-principle half.} With $K_\perp^a=\D_\perp^a\Phi_A$, and
$\chi$ increasing backwards along the ray so that $\dd t = -a\,\dd\chi$,
Eq.~(\ref{eq:de}) integrated forwards to reception gives the rotation of the
direction, relative to the threading, accumulated along the ray. Throughout,
$\Delta e_\perp$ is the change in direction between the point in question and
reception:
\begin{equation}
  \Delta e_\perp\big|_{K} = -\int_0^{\chi_*}\!\nabla_{\!\perp}\Phi_A\,\dd\chi .
  \label{eq:bendK}
\end{equation}
This is the bending measured by observers who are accelerated to remain on the
threading. For a point mass it is $2GM/b$, half the total.

\subsection{Frame invariance at the multipoles, and the threading shift}
\label{sec:frameinv}

\emph{Invariance.} The deflection depends only on $\Phi_A-\Phi_H$, and its
trace-free screen-projected Hessian is the electric part of the Weyl tensor:
$E_{ab}\approx\frac12\D_{\la a}\D_{b\ra}(\Phi_A-\Phi_H)$ is Eq.~(24) of
Ref.~\cite{GebbieDunsbyEllis1999}, and $E_{ab}$ vanishes in the background. By
the Stewart--Walker lemma the deflection is therefore gauge-invariant: it does
not depend on the map to the background. That is not yet independence of the
physical congruence, which is a separate statement, and the stronger one is in
the same place. Below its Eq.~(35), Ref.~\cite{GebbieDunsbyEllis1999} records that
$\rho$, $p$, $\pi_{ab}$, $E_{ab}$, $H_{ab}$ and the temperature anisotropies
$\tau_{A_\ell}$ for $\ell>1$ are all unchanged under $\tilde u_a\simeq u_a+v_a$,
only the dipole moving, $\tilde\tau_a\approx\tau_a-v_a$, its Eq.~(37) --- frame
invariance at small tilt, which is what is needed here --- at linear order in
$v$, and for multipoles read against a common direction label. Frame
covariance is thus asserted directly at the multipoles and not only at the
deflection, which sets the standard the rest of this section has to meet: with
the emitter and observer frames held fixed, any dependence on the
\emph{intermediate} threading surviving at $\ell\ge2$ is an artefact of the
reduction and not physics. What the endpoints themselves carry is not an
artefact and is not covered by that statement: it is the Doppler and aberration
transformation between fixed physical frames of Sec.~\ref{sec:relabel}, which
at reception carries the rank. Each threading divides the
deflection differently between the aberration operator and the geometry of its
sections; the sum is the same.
Sections~\ref{sec:newtonian-pair} and~\ref{sec:cdm-triple} compute the aberration
side of each division term by term and find one operator in both; that the same
operator in two threadings gives the same observed sky rests in addition on
Eq.~(\ref{eq:dK}). The side carried by the sections moves too --- by the
$-\ts\frac13\Theta\sigma_{ab}$ of Eq.~(\ref{eq:gauss-tf}), which is first order as
soon as the threading has shear --- and it is reduced here once only, in the
shear-free threading of Sec.~\ref{sec:curvhalf}. It does not have to be reduced
twice. The two skies are brought together in Sec.~\ref{sec:relabel} by a
pointwise relabelling of the direction argument, a statement about the photon's
two labels that holds whatever division either threading makes.

In the metric second-order hierarchy the same two pieces appear as the
observer-acceleration and spatial-connection parts of the photon direction
equation, and combine in a single term \cite{BenekeFidler2010}. In the $1+3$
variables they sit in different terms of Eq.~(\ref{eq:master}): the explicit
source and the free-streaming operator.

The shear, vorticity and trace parts of $\D_bv_a$ in
Eq.~(\ref{eq:frame-transf}) recombine into $e^b\D_bv_a$, so that
$\tilde K_\perp^a - K_\perp^a \simeq (\dot v^a + Hv^a + e^b\D_bv^a)_\perp$,
which is Eq.~(\ref{eq:dK}). Directions in the two frames are related at each
event by aberration, $\tilde e^a = e^a - v_\perp^a$, and energies by Doppler,
$\tilde E = E(1-v_ae^a)$. The $Hv$ term appears because each frame transports
its own directions against its own expansion, and $\tilde u^a$ projects onto
the rest space of $u^a$ as $v^a$; App.~\ref{app:shift} derives
Eq.~(\ref{eq:dK}) a second time from the photon's direction equation alone,
where that term arises from the difference of the two screen projectors and
carries the redshift rate as its coefficient. Substituting
$\tau(x,e)\to\tau(x,\tilde e+v_\perp)-v_ae^a$ into the $u$-threaded equation
and applying the chain rule generates $(\dd v/\dd t+Hv)_\perp\cdot\partial_e\tau$,
which added to $K_\perp\cdot\partial_e\tau$ gives
$\tilde K_\perp\cdot\partial_e\tau$. The two hierarchies are therefore one
equation in two coordinate systems on photon phase space;
Sec.~\ref{sec:relabel} carries the change of coordinates out on the multipoles,
where it is algebraic. The closure needs the
sign of Sec.~\ref{sec:shearsign}: Eq.~(\ref{eq:frame-transf}) mixes the shear and
acceleration couplings, and the mixture closes on Eq.~(\ref{eq:dK}) only if
both are the one operator $K_\perp\cdot\partial_e$.

The boost enters through its gradient as well as its magnitude. For a
sub-horizon mode $v\sim(H/k)\Delta$ is small, but
$e^b\D_bv\sim kv\sim H\Delta\sim\sigma$ is not. So a small boost leaves
$\tau_{A_\ell}$, $\ell>1$, invariant for linear anisotropies, but at second
order it acts through $v_\perp\cdot\partial_e\tau$, whose multipole
coefficients are $O(\ell)$, the order of the couplings themselves.

\subsection{The Newtonian threading: the acceleration pair}
\label{sec:newtonian-pair}

In the Newtonian threading $\tilde\sigma_{ab}=0=\tilde\omega_a$ and
$\tilde A_a=\D_a\Phi_A$, so $K^a=\D^a\Phi_A$ carries no direction dependence of
its own. Equation~(\ref{eq:abersplit}) then has only two structures to make.
The transverse sum needs no reduction and contributes at rank $\ell$ from
$n=\ell+1$ alone; the longitudinal sum reduces by Eq.~(\ref{eq:red-A}) and
contributes from $n=\ell-1$ and $n=\ell+1$. Collecting, the $O(\ell)$ source in
this threading is
\begin{equation}
  \bigl(\delta\dot\tau\bigr)^{A_\ell}_{\!_{NL}} =
  -(\ell-1)\,A^{\la a_\ell}\tau^{A_{\ell-1}\ra}
  + \frac{(\ell+1)(\ell+2)}{2\ell+3}\,A_b\tau^{bA_\ell} ,
  \label{eq:newt-pair}
\end{equation}
which is the aberration column of Table~\ref{tab:coefficients}, rows one and
two, transposed; at large $\ell$ it is
$\ell[-A^{\la a_\ell}\tau^{A_{\ell-1}\ra}+\frac12A_b\tau^{bA_\ell}]$, the
acceleration pair of Eq.~(\ref{eq:nl-source}).

The division is then visible term by term. The $\ell-1$ coupling comes entirely
from the longitudinal subtraction --- it is the row with $r=1$ and $t=0$ --- and
the $\ell+1$ coupling from both pieces at once. They are the rank-raising and
rank-lowering images of one vector field on the sphere, $\D_\perp\Phi_A$, and
not two couplings that happen to appear together. Their opposite signs are
forced by that, and by the transposition argument of Sec.~\ref{sec:shearsign} a
common sign would make the pair a growth rate proportional to $\ell$ rather than
a remapping. Nothing else appears: with $\sigma_{ab}=0=\omega_a$ there are no
other rows.

\subsection{The line-of-sight integral, the distance factor and the efficiency}
\label{sec:efficiency}

\begin{figure}[!htbp]
\includegraphics[width=\columnwidth]{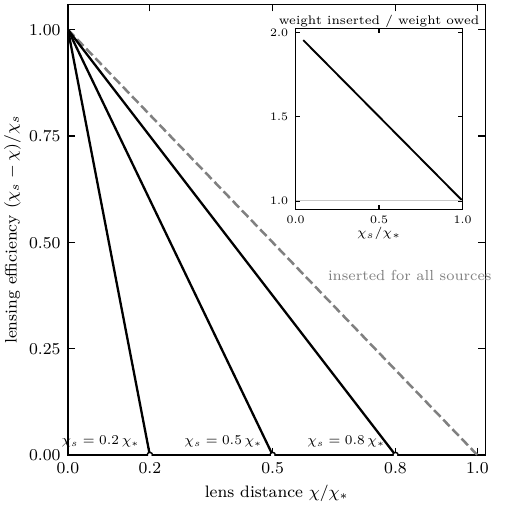}
\caption{The lensing efficiency: one kernel per source. The direction derivative
at $\chi$ carries the distance to its own source [Eq.~(\ref{eq:dtau})], so a
lens at $\chi$ weights a source at $\chi_s$ by $(\chi_s-\chi)/\chi_s$ --- the
primary with $\chi_s=\chi_*$, and each linear secondary with its own. The
efficiency is not inserted; it follows from that distance factor, which is the one
place the covariant route produces a kernel the remapping treatment has to
supply. A single deflection field applied to the whole sky necessarily lenses
every source with the recombination kernel (dashed). \emph{Inset:} the two
\emph{integrated} kernels compared, exactly. A source receives total weight
$\chi_s/2$ under its own kernel and $\chi_s(1-\chi_s/2\chi_*)$ under the
recombination one, a ratio of $2-\chi_s/\chi_*$, so a source halfway to last
scattering is over-weighted by half. That is a statement about the bare
geometry and not the fractional error on any particular secondary: for a lens
at $\chi_l$ the ratio is
$(1-\chi_l/\chi_*)/(1-\chi_l/\chi_s)$, and what a given secondary sees depends
on where its lenses sit along the ray. The kernel is geometry in the variable plotted: nothing here depends on a
cosmological parameter. The distance $\chi_*$ that sets the scale does, and the
figure is drawn against $\chi/\chi_*$ for that reason.}
\label{fig:efficiency}
\end{figure}

\emph{The efficiency.} The weakly nonlinear solution carries the line-of-sight
integral of $K_\perp\cdot\partial_e\tau^{(1)}$ on the background ray. At
$x(\chi)$ the free-streamed first-order field is
$\tau^{(1)}(x,e)=\tau_E\bigl(x-(\chi_*-\chi)e,\,e\bigr)$, plus the linear
secondaries between $x$ and the emission point, so its direction derivative is
\begin{equation}
  \partial_{e}\tau^{(1)}\big|_{x(\chi)}
  = -(\chi_*-\chi)\,\nabla_{\!\perp}\tau_E
  + \partial_e\tau_E\big|_{x_E} + \ldots .
  \label{eq:dtau}
\end{equation}
The first part is the spatial pattern of the emitted field seen at distance
$\chi_*-\chi$; the second is its intrinsic direction dependence. Substituting,
\begin{widetext}
\begin{equation}
  \int_0^{\chi_*}\!K_\perp\cdot\partial_e\tau^{(1)}\,a\,\dd\chi
  = \underbrace{\Bigl[-\!\int_0^{\chi_*}\!(\chi_*-\chi)\,\nabla_{\!\perp}\Phi_A\,\dd\chi\Bigr]
    \cdot\nabla_{\!\perp}\tau_E}_{\text{lensing, }\Phi_A\text{ part}}
  \;-\;\underbrace{\Delta e_\perp\big|_K\cdot\partial_e\tau_E}_{\text{aberration at }E}
  \;+\;(\text{lensing of secondaries}) .
  \label{eq:Kint}
\end{equation}
\end{widetext}
The geodesic map displaces the emission point by
\begin{equation}
  \delta x_\perp = -\int_0^{\chi_*}\!(\chi_*-\chi)\,
  \nabla_{\!\perp}(\Phi_A-\Phi_H)\,\dd\chi ,
  \label{eq:deltax}
\end{equation}
so that $\tau_R$ contains a term $\delta x_\perp\cdot\nabla_{\!\perp}\tau_E$. The first term
of Eq.~(\ref{eq:Kint}) is exactly its $\Phi_A$ part, in sign, weight and
normalisation, and the section term of Eq.~(\ref{eq:bendH}), propagated the same
way, supplies the $\Phi_H$ part. The lensing efficiency $(\chi_*-\chi)/\chi_*$
is not inserted. It comes from the distance factor in Eq.~(\ref{eq:dtau})
(Fig.~\ref{fig:efficiency}). The
unweighted bending enters only the aberration at emission, whose coupling is
$O(1)$. The linear secondaries are lensed with their own efficiency,
$(\chi'-\chi)/\chi'$ for a source at $\chi'$.

\subsection{The spatial-curvature half}
\label{sec:curvhalf}

The multipoles are tensors on the rest space, and the free-streaming term of
the hierarchy differentiates them with $\D_b$, the derivative of the sections
themselves. Expanding about the background replaces $\D_b$ by the background
derivative $\mathring\D_b$ together with the difference of the two connections,
$C^c{}_{ab}\equiv\Gamma^c{}_{ab}-\mathring\Gamma^c{}_{ab}$, which is a tensor and is
symmetric in its lower pair:
\begin{equation}
  \D_b\Pi^{A_\ell} = \mathring\D_b\Pi^{A_\ell}
    + \sum_{i=1}^{\ell}C^{a_i}{}_{bc}\,\Pi^{a_1\cdots c\cdots a_\ell} .
  \label{eq:conndiff}
\end{equation}
There is one term for each index, each first order in the geometry of the
sections, so on a first-order $\Pi_{A_\ell}$ the sum is $O(\ep^2\ell)$. It sits
inside the free-streaming operator of Eq.~(\ref{eq:master}) and not in
Eq.~(\ref{eq:nl-source}), and it is exactly what linearising the free-streaming
term discards.

The sections of the Newtonian threading are shear-free and irrotational and
their scalar geometry is conformally flat at first order, so the connection
difference is that of a conformal rescaling by a scalar $\Phi_C$ --- subscripted
for the connection difference $C^c{}_{ab}$ it generates, and carried as an
undetermined scalar because Eq.~(\ref{eq:PhiCPhiH}) below derives what it is
rather than assuming it ---
\begin{equation}
  C^c{}_{ab} = \delta^c_a\D_b\Phi_C + \delta^c_b\D_a\Phi_C - h_{ab}\D^c\Phi_C .
  \label{eq:conformalC}
\end{equation}
Write $g_a\equiv\D_a\Phi_C$. Contracting Eq.~(\ref{eq:conformalC}) with $e^b$,
\begin{equation}
  e^bC^a{}_{bc} = e^ag_c + \delta^a_c\,(g_be^b) - e_c\,g^a ,
  \label{eq:eC}
\end{equation}
and the three terms reduce in turn. All $\ell$ summands of
Eq.~(\ref{eq:conndiff}) are equal once contracted with the symmetric
$O_{A_\ell}$, so each carries a factor $\ell$.

The middle term of Eq.~(\ref{eq:eC}) gives $+\ell\,(g_be^b)\Pi_{A_\ell}O^{A_\ell}$.
The last gives $-\ell\,g^a\,(e_c\Pi^{c}{}_{A_{\ell-1}})O_{aA_{\ell-1}}$, and
inverting Eq.~(\ref{eq:eO}) to $O_a{}^{A_{\ell-1}} = e_aO^{A_{\ell-1}} -
\frac{\ell-1}{2\ell-1}h_a{}^{\la a_{\ell-1}}O^{A_{\ell-2}\ra}$ splits it into a
part $-\ell\,(g_be^b)\Pi_{A_\ell}O^{A_\ell}$, which cancels the middle term
exactly, and a remainder. The first term lowers one basis index by C6.1. What
is left, per source rank $\ell$, is
\begin{equation}
  \frac{\ell^2}{2\ell-1}\,g_c\Pi^{c}{}_{A_{\ell-1}}O^{A_{\ell-1}}
  + \frac{\ell(\ell-1)}{2\ell-1}\,g^a\,e^c\Pi_{caA_{\ell-2}}O^{A_{\ell-2}} .
  \label{eq:conn-two}
\end{equation}
Both are direction derivatives. By Eq.~(\ref{eq:dPide}) the first is
$\frac{\ell}{2\ell-1}g^a\partial_a\Pi$; the second is
$\frac{1}{2\ell-1}g^a\,e^c\partial_c\partial_a\Pi$, and since $\partial_a\Pi$ is
homogeneous of degree $\ell-1$ in $e^a$, Euler's identity turns $e^c\partial_c$
on it into a factor $\ell-1$. The two coefficients sum to
$(\ell+\ell-1)/(2\ell-1)=1$, and therefore
\begin{equation}
  \delta\bigl(e^b\D_b\Pi\bigr) = (\D^a\Phi_C)\,\frac{\partial\Pi}{\partial e^a} ,
  \label{eq:curvop}
\end{equation}
exactly, at every $\ell$, with no coefficient of its own.

What Eq.~(\ref{eq:curvop}) is, stated plainly. Compared with
Eq.~(\ref{eq:abersplit}), free streaming on the perturbed sections \emph{is} an
aberration operator, with $K_\perp^a$ replaced by $\D_\perp^a\Phi_C$. The
spatial-curvature half is not a second mechanism running alongside the
equivalence-principle half; it is the same operator reached by a different
route, which is why the two halves are equal for a point mass and why neither
needs its own distance factor, its own efficiency or its own reduction. Everything
derived in Sec.~\ref{sec:lensing} carries over unchanged.

Two things follow. First, free streaming sits on the left of the hierarchy and
the aberration term on the right, so Eq.~(\ref{eq:curvop}) acts with the
opposite sign and
\begin{equation}
  \Delta e_\perp\big|_{\rm sections} = +\int_0^{\chi_*}\!\nabla_{\!\perp}\Phi_C\,\dd\chi ,
  \label{eq:bendH}
\end{equation}
which is the second half of the deflection. Second, $\D^a\Phi_C$ in
Eq.~(\ref{eq:curvop}) is not transverse. Splitting it,
$(\D^a\Phi_C)\partial_a = (\D_\perp^a\Phi_C)\partial_a + (e^a\D_a\Phi_C)\,e^b\partial_b$,
and the second piece returns the rank rather than a deflection: it is a
magnification along the ray, not a remapping of directions, and it is not part
of the bending.

\subsubsection{The conformal factor}
\label{sec:conformal-factor}

Equation~(\ref{eq:curvop}) is complete except for $\Phi_C$, and $\Phi_C$ is
fixed by the intrinsic curvature of the sections rather than by anything in the
photon problem. For a first-order conformal difference the intrinsic Ricci
tensor of a three-dimensional space is $-\D_a\D_b\Phi_C-h_{ab}\D^2\Phi_C$
[Table~\ref{tab:conventions}, C2.4], whose trace-free part is
\begin{equation}
  {}^{(3)}\!R_{\la ab\ra} = -\D_{\la a}\D_{b\ra}\Phi_C .
  \label{eq:3ricci}
\end{equation}
So $\Phi_C$ is a potential for the trace-free intrinsic curvature of the
sections, and for nothing else.

The other potential is already defined that way. For scalar perturbations
Ref.~\cite{GebbieDunsbyEllis1999} writes the electric Weyl tensor and the
anisotropic stress in potentials, its Eqs.~(24) and~(25), and combines them as
its Eq.~(26),
\begin{equation}
  E_{ab}-\ts\frac12\pi_{ab}\approx\D_{\la a}\D_{b\ra}\Phi_A ,
  \qquad
  E_{ab}+\ts\frac12\pi_{ab}\approx-\D_{\la a}\D_{b\ra}\Phi_H ,
  \label{eq:potentials}
\end{equation}
so that $\Phi_H$ is a potential for $-(E_{ab}+\frac12\pi_{ab})$ in exactly the
sense that $\Phi_C$ is one for $-{}^{(3)}\!R_{\la ab\ra}$. The first of
Eqs.~(\ref{eq:potentials}) is where $\Phi_A$ comes from here too: in the
shear-free frame the shear propagation equation reduces to
$\D_{\la a}A_{b\ra}\approx E_{ab}-\frac12\pi_{ab}$, hence
$A_a\approx\D_a\Phi_A$, which is Eq.~(49) of the same paper and the line the
Newtonian threading of Sec.~\ref{sec:threading} rests on.

The two potentials therefore agree when the trace-free intrinsic curvature of
the sections is $E_{ab}+\frac12\pi_{ab}$, and it is. The Gauss embedding
equation gives the Ricci tensor of the orthogonal $3$-spaces of a vorticity-free
congruence as Eq.~(54) of Ref.~\cite{EllisVanElst1998}, whose trace-free part is
\begin{equation}
  {}^{(3)}\!R_{\la ab\ra} = -\dot\sigma_{\la ab\ra} - \Theta\sigma_{ab}
    + \D_{\la a}A_{b\ra} + A_{\la a}A_{b\ra} + \pi_{ab} ,
  \label{eq:gauss-54}
\end{equation}
and the shear propagation equation, Eq.~(31) of the same reference, gives
$\dot\sigma_{\la ab\ra} - \D_{\la a}A_{b\ra}$ as
$-\frac23\Theta\sigma_{ab} + A_{\la a}A_{b\ra} - \sigma_{\la a}{}^c\sigma_{b\ra c}
- (E_{ab}-\frac12\pi_{ab})$ when $\omega_a=0$. Substituting, the acceleration
terms cancel identically and
\begin{equation}
  {}^{(3)}\!R_{\la ab\ra} = E_{ab} + \ts\frac12\pi_{ab}
    - \ts\frac13\Theta\sigma_{ab} + \sigma_{\la a}{}^c\sigma_{b\ra c} .
  \label{eq:gauss-tf}
\end{equation}
Every correction carries a shear, so for the shear-free threading
Eq.~(\ref{eq:gauss-tf}) is exact, not linearised
[Table~\ref{tab:conventions}, C2.7]. With
Eq.~(\ref{eq:potentials}) it gives
\begin{equation}
  \D_{\la a}\D_{b\ra}\Phi_C = \D_{\la a}\D_{b\ra}\Phi_H ,
  \qquad\text{so}\qquad \Phi_C = \Phi_H ,
  \label{eq:PhiCPhiH}
\end{equation}
on every mode of nonzero wavenumber. The second step is not free: the kernel
of $\D_{\la a}\D_{b\ra}$ on a flat patch contains the constants, the linear
functions and the isotropic quadratic $x^cx_c$. Those are precisely the parts
that carry no lensing --- a constant, a uniform deflection of the whole sky and
a uniform convergence --- so fixing them is a choice of boundary condition and
not a physical input, and the same remark applies to reading
$A_a\approx\D_a\Phi_A$ off equality of PSTF derivatives.
This is the identification Eq.~(\ref{eq:bendH}) was written with, and with it
the two halves are equal for a point mass. It also settles the remark of
Ref.~\cite{GebbieDunsbyEllis1999} that $\Phi_A$ is the Newtonian potential of
the shear-free frame while $\Phi_H$ ``can be interpreted as a curvature
perturbation'': here the second reading is not an interpretation but the
definition the deflection uses. The $\Phi_A$ side rests on two sources in the
same way: Eq.~(49) of Ref.~\cite{GebbieDunsbyEllis1999} and Eq.~(31) of
Ref.~\cite{EllisVanElst1998} at $\sigma_{ab}=0$ both give
$A_a\approx\D_a\Phi_A$ [Table~\ref{tab:conventions}, C2.6].

With $\Phi_C=\Phi_H$, the sum of Eqs.~(\ref{eq:bendK}) and (\ref{eq:bendH}) is
the total bending (Fig.~\ref{fig:halves}),
$-\int\nabla_{\!\perp}(\Phi_A-\Phi_H)\,\dd\chi$, which for $\Phi_H=-\Phi_A$ is
twice Eq.~(\ref{eq:bendK}) and $4GM/b$ for a point mass.\footnote{%
This half is absent if the free-streaming term is linearised and the explicit
couplings of Eq.~(\ref{eq:nl-source}) are taken as the whole
$O(\ep^2\ell)$ source.}

\subsection{Bulk and endpoint}
\label{sec:bulkendpoint}

The local source of Eq.~(\ref{eq:nl-source}) is $O(\ell)$, while the aligned
limit of Sec.~\ref{sec:aligned} finds its two brackets cancelling at leading
order. Read side by side those look contradictory. They are not, and the
resolution is one integration by parts along the ray.

Write $\Delta e_\perp(\chi)=-\int_0^\chi\!K_\perp\,a\,\dd\chi'$ for the
deflection accumulated between the observer and $\chi$, so that
$\dd\Delta e_\perp/\dd\chi=-aK_\perp$ and $\Delta e_\perp(0)=0$. Then
\begin{align}
  \int_0^{\chi_*}\!\! aK_\perp\!\cdot\!\partial_e\tau^{(1)}\,\dd\chi
  &= -\Bigl[\Delta e_\perp\!\cdot\!\partial_e\tau^{(1)}\Bigr]_{\chi_*}
  \nonumber\\
  &\quad + \int_0^{\chi_*}\!\!\Delta e_\perp\!\cdot\!\frac{\dd}{\dd\chi}
    \bigl(\partial_e\tau^{(1)}\bigr)\dd\chi .
  \label{eq:byparts}
\end{align}
The endpoint term is $\Delta e_\perp(\chi_*)$ against the intrinsic direction
dependence of the emitted field. It is the aberration at emission of
Eq.~(\ref{eq:Kint}) and it is $O(1)$, because $\partial_e\tau_E$ at fixed
position carries $\ell\le2$ in the baryon frame. The bulk term is where the
$\ell$ survives: by Eq.~(\ref{eq:dtau}) the only $\chi$-dependence of
$\partial_e\tau^{(1)}$ is the distance factor it carries, so
$\dd(\partial_e\tau^{(1)})/\dd\chi=+\nabla_{\!\perp}\tau_E$ and the bulk is
$\bigl[\int_0^{\chi_*}\Delta e_\perp\dd\chi\bigr]\cdot\nabla_{\!\perp}\tau_E$.
Exchanging the order of the double integral,
\begin{equation}
  \int_0^{\chi_*}\!\!\Delta e_\perp(\chi)\,\dd\chi
  = -\int_0^{\chi_*}\!\!(\chi_*-\chi)\,\nabla_{\!\perp}\Phi_A\,\dd\chi ,
  \label{eq:swap}
\end{equation}
which is the $\Phi_A$ part of Eq.~(\ref{eq:deltax}). The two terms of
Eq.~(\ref{eq:byparts}) are the two terms of Eq.~(\ref{eq:Kint}), and the
efficiency $(\chi_*-\chi)/\chi_*$ comes out of the exchange of integrals rather
than being put in.

So, in three sentences. The local operator is not a source but the generator of
a remapping of directions, and a generator integrated along the ray composes
into one finite remapping instead of accumulating. A remapping does nothing at
leading order to a spectrum that is smooth in $\ell$ at fixed source position,
which is the cancellation of Sec.~\ref{sec:aligned}. What survives is the one
place where the remapping meets a transverse gradient --- the distance factor,
which
is the lensing of Sec.~\ref{sec:lensing}, and which is $O(\ell)$ because
$\nabla_{\!\perp}$ acting on a pattern of angular scale $1/\ell$ is.

\subsubsection{\texorpdfstring{The longitudinal term of
Eq.~(\ref{eq:curvop})}{The longitudinal term}}

The same reading disposes of the one term Sec.~\ref{sec:curvhalf} left over. The
direction $e^a$ of the hierarchy is a unit vector in the metric of the sections
[Table~\ref{tab:conventions}, C1.5], so under the conformal difference of
Eq.~(\ref{eq:conformalC}) it is $e^a=e^{-\Phi_C}\mathring e^a$ with
$\mathring e^a$ the unit vector of the background. Trace-freeness of a symmetric
tensor with upper indices is the same condition for conformally related metrics,
so the basis carries the factor and the components carry its inverse,
\begin{equation}
  O^{A_\ell}=e^{-\ell\Phi_C}\,\mathring O^{A_\ell} ,
  \qquad
  \Pi_{A_\ell}=e^{+\ell\Phi_C}\,\mathring\Pi_{A_\ell} ,
  \label{eq:basisfactor}
\end{equation}
the second because $\Pi$ itself does not depend on the basis. That factor is the
whole of the longitudinal term: differentiating it along the ray gives
$\ell\,(\dd\Phi_C/\dd\chi)$ acting on $\Pi_{A_\ell}$, whose spatial part is the
piece $(e^a\D_a\Phi_C)\,e^b\partial_b\Pi$ of Eq.~(\ref{eq:curvop}) and whose
time part is the matching term in $\dot\Pi$. It is therefore not a coupling but
the derivative of a normalisation, and it is absent from the hierarchy written
for $\mathring\Pi_{A_\ell}$ --- multipoles labelled by the directions of the
background, which is what an observer's harmonics supply. The transverse term of
Eq.~(\ref{eq:curvop}) is unchanged by the rescaling to this order, the two
halves of the deflection are unaffected, and the enumeration of
Table~\ref{tab:physics} stands as written. What the factor does record is real
but is not an $O(\ell)$ coupling: at each end of the ray it says that the
angular size of a fixed physical structure is measured against the local
geometry and not the background's.

Zibin and Scott \cite{ZibinScott2008} show, exactly, that the observed
temperature in direction $-n^\mu$ is $T_R/T_E = \exp\int H_nN\,\dd t$, with
$H_n = \frac13\theta + \sigma_{\mu\nu}n^\mu n^\nu + a_\mu n^\mu$ (their
Eqs.~(16)--(24)). The integral is independent of the congruence between the
endpoints once the observer's four-velocity at $R$ and the emitter's frame at
$E$ are fixed; at linear order the integrand changes by a total derivative that
the endpoint terms absorb (their Eqs.~(B8)--(B13)). $H_n$ is the redshift term
of Eq.~(\ref{eq:liouville}), and $\mathcal B$ of Eq.~(\ref{eq:B}) is its $1+3$
form. Their argument needs only the energy half of $u_{a;b}p^b$, because at
linear order the direction multiplies an isotropic background. The $O(\ell)$
couplings are the first place where the direction half matters, and the same
argument applies to it: the direction seen at $R$ is $p^a_R$ projected into the
rest space of $u_R^a$, and the emission point is fixed by the null geodesic.
Equation~(\ref{eq:dK}) is the direction counterpart of their Eq.~(B9).

A zero-shear frame is fixed only up to a tilt $v^a$ with
$\D_{\la a}v_{b\ra}=0$ (their Sec.~VI), and by Eq.~(\ref{eq:dK}) the tilt
changes the $O(\ell)$ source. Evaluated at $R$ it aberrates the observed sky,
which beyond first order moves power into every multipole (their App.~C). A
$C_\ell$ stated at $O(\ep^2\ell)$ in a zero-shear threading, and not
transformed to the observer's frame, is defined only up to this aberration.

\subsection{The enumeration}
\label{sec:residue}

A term-by-term enumeration of the geodesic map at $O(\ep^2)$
finds no other $O(\ep^2\ell)$ content: lensing of the primary and of the
secondaries, lensing by tensors, and observer aberration exhaust it. The
enumeration is Table~\ref{tab:enumeration}.

What makes it an enumeration rather than a list is that the rows are counted
and not collected. Expanding the observed temperature,
$\tau_R(e_R)=\tau_E(x_E,e_E)+\int(\text{redshift})$, to $O(\ep^2)$ along the
true geodesic leaves every term a product of two $O(\ep)$ factors, one of which
perturbs the map or the observer's frame. The perturbation of the map at the
emission point has six components --- two transverse and one radial of
position, two of direction and one of energy --- the observer's frame has
three, and at this order nothing else can multiply a first-order field. Of
those nine, four put a transverse or angular gradient on the high-$\ell$
pattern of $\tau$: the two of $\delta x_\perp$ and the two of $v_{R\perp}$, the
transverse displacement of the emission point and the observer's own transverse
velocity. The other five do not, for the reasons Table~\ref{tab:enumeration}
gives in each case, and the second-order parts of the map meet only $\bar\tau$,
which is homogeneous and isotropic.

Two things lie outside the count rather than inside it, and both are carried
elsewhere. The second-order emitted field is not a perturbation of the map but
part of what the map acts on; in the baryon frame tight coupling holds it at
$\ell\le2$, so no angular gradient meets a high-$\ell$ pattern in it. And what
happens to the photons between the endpoints enters Eq.~(\ref{eq:master}) as a
source, not as a coupling to $\tau_{A_\ell}$; its one part beyond first order is
the Thomson term of Eq.~(\ref{eq:hierarchy-exact}), counted in $\vb$ and not in
$\ep$ [Table~\ref{tab:physics}, last row].

Because the count is over the geometry of
the map and not over what sources it, no mode type can be missed: the scalar,
vector and tensor contributions enter the same six components.

A treatment that lenses the total sky with the recombination efficiency
approximates the separate efficiency of the secondaries; that is a matter of
implementation. The $O(\ep^2\ell^0)$ remainder --- the redshift parts of
Table~\ref{tab:coefficients} and the $5\ell/(2\ell+3)$ term --- is
frame-covariant in combination by the exact result of
Ref.~\cite{ZibinScott2008}, and belongs to the ordinary Rees--Sciama and
integrated Sachs--Wolfe terms.

\section{Frame covariance}
\label{sec:frame}
\begin{figure}[!htbp]
\includegraphics[width=\columnwidth]{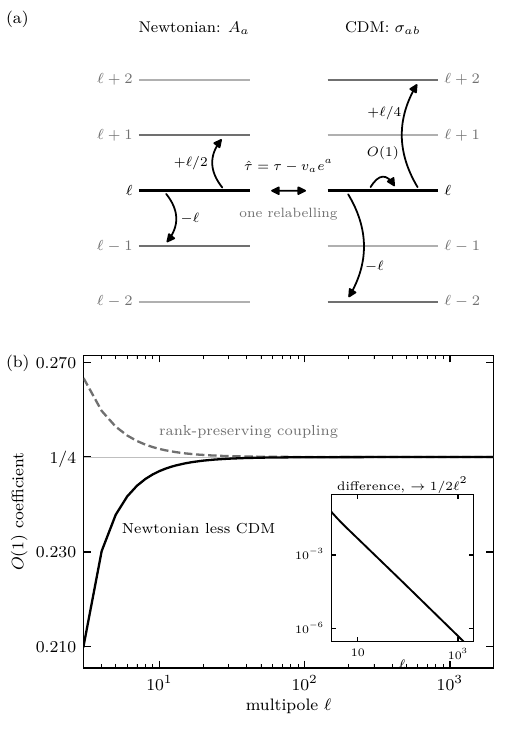}
\caption{The two threadings. \emph{(a)} The same $O(\ep^2\ell)$ source in each:
the Newtonian threading carries it on the acceleration, reaching one rank
either side, and the CDM threading on the shear, reaching two and carrying in
addition a coupling that preserves rank. The coefficients are the large-$\ell$
limits of Eqs.~(\ref{eq:newt-pair}) and~(\ref{eq:cdm-triple}); the ladder
positions are drawing coordinates. The two patterns are one operator in two
labels, and the map between them is the pointwise relabelling of
Eq.~(\ref{eq:relabel}), generated by Eq.~(\ref{eq:dK}). \emph{(b)} Where the
$O(1)$ parts close. The amount by which the CDM bracket falls short of the
Newtonian one, Eq.~(\ref{eq:resN}) less Eq.~(\ref{eq:resC}), against the
rank-preserving coupling of Eq.~(\ref{eq:rankpres}) --- the rung that only the
CDM threading has: both tend to $\frac14$, and from opposite sides, so the
quarter the CDM threading is missing is the coupling the Newtonian threading
does not have. All three are evaluated in exact rationals. \emph{Inset:} what
is left of the difference, falling as $1/2\ell^2$ --- the two brackets are
stencils of width two and width four and carry different curvature content,
which is the first order at which they can differ and the first the
decomposition does not resolve.}
\label{fig:threadings}
\end{figure}

The threading is fixed inside the equations, not before them. Changing it
changes which
kinematic quantity carries the explicit coupling --- the acceleration in one
case, the shear in the other --- without changing the observed sky, because the
emitter and observer frames are fixed independently of the threading between
them. This section exhibits that: the generator that relates the two, the
division computed in each threading, the relabelling that undoes the difference
between them on the multipoles, and the aligned configuration in which the two
forms can be written side by side.

\subsection{The CDM threading: the shear triple, term by term}
\label{sec:cdm-triple}

The two divisions are drawn side by side in Fig.~\ref{fig:threadings}(a).

In the CDM threading $\tilde A_a=0$ and, by Eq.~(\ref{eq:twoframes}),
$\sigma^{\!_C}_{ab}=\D_{\la a}v^{\!_C}_{b\ra}$, so
$K^a=\sigma^{\!_C\,a}{}_{b}e^b$. This $K^a$ \emph{does} depend on the direction,
which is why it reaches $\ell\pm2$ where the acceleration reached $\ell\pm1$.
For scalar modes $v^{\!_C}_a$ is a gradient, so $\mbox{curl}\,v^{\!_C}=0$ and
\begin{equation}
  K^a_{\!_C} = \bigl(\D^{\la a}v^{b\ra}_{\!_C}\bigr)e_b
  = \bigl(e^b\D_bv^a_{\!_C}\bigr)_{\!\perp} + (\text{longitudinal}) ,
  \label{eq:Kcdm}
\end{equation}
the transverse rate at which the boost to the Newtonian threading changes along
the ray. That is the same object Eq.~(\ref{eq:dK}) produces, reached from the
kinematics rather than from the frame shift.

The reduction runs as before. The transverse sum of
Eq.~(\ref{eq:abersplit}) now needs Eq.~(\ref{eq:red-sT}) and contributes at rank
$\ell$ from $n=\ell$ and $n=\ell+2$; the longitudinal sum needs
Eq.~(\ref{eq:red-s}) and contributes from $n=\ell-2$, $n=\ell$ and $n=\ell+2$.
Collecting,
\begin{widetext}
\begin{equation}
  \bigl(\delta\dot\tau\bigr)^{A_\ell}_{\!_{NL}} =
  -(\ell-2)\,\sigma^{\la a_\ell a_{\ell-1}}\tau^{A_{\ell-2}\ra}
  + (\ell+3)\,c_2\,\sigma_{bc}\tau^{bcA_\ell}
  + \frac{3\ell}{2\ell+3}\,\sigma_b{}^{\la a_\ell}\tau^{A_{\ell-1}\ra b} ,
  \label{eq:cdm-triple}
\end{equation}
\end{widetext}
rows three, four and five of Table~\ref{tab:coefficients}, transposed. At large
$\ell$ the first two are
$\ell[-\sigma^{\la a_\ell a_{\ell-1}}\tau^{A_{\ell-2}\ra}
+\frac14\sigma_{bc}\tau^{bcA_\ell}]$, the shear pair of
Eq.~(\ref{eq:nl-source}), and the third is $O(1)$ and is not part of the
$O(\ell)$ source.

Term by term, then, the division is the same one. In the Newtonian threading
the whole $O(\ell)$ source is the acceleration pair; in the CDM threading the
whole of it is the shear pair; and both are the PSTF image of the one operator
$K_\perp^a\partial/\partial e^a$, differing only in which kinematic quantity
supplies $K^a$ and therefore in how many ranks the reduction reaches. The
$\ell-2$ coupling is again the row with $r=1$ and $t=0$, the $\ell+2$ coupling
again draws on both pieces, and their signs are again opposite for the same
reason. The one structural difference between the two threadings is the
rank-preserving coupling, which exists only when $K^a$ depends on direction and
which is $O(1)$ in every multipole.

Equations~(\ref{eq:newt-pair}) and~(\ref{eq:cdm-triple}) show that the two
threadings carry the same operator. That they therefore give the same observed
sky rests on Eq.~(\ref{eq:dK}), which App.~\ref{app:shift} derives a second
time from the photon's own direction equation, sharing no step with the
derivation in Sec.~\ref{sec:threading}. What that leaves is the statement that
the difference between the two divisions accumulates to nothing, and that is
the subject of Sec.~\ref{sec:relabel}.

\subsection{The relabelling, and the difference as an endpoint term}
\label{sec:relabel}

The difference is not merely a total derivative along the ray. It is a
pointwise algebraic relabelling, so that once the emitter and observer frames
are named the two threadings differ only by the transformation between those
frames --- at every $\ell$, and not only at $O(\ep^2\ell)$.

\emph{One label for both.} A single photon is split by both threadings: its
energies are related by Eq.~(\ref{eq:doppler}) and its directions by
Eq.~(\ref{eq:aber}), which for this pair read
$E_{\!_C}=E(1-v^{\!_C}_ae^a)$ and
$e^a_{\!_C}=e^a-v^a_{\!_C}+(v^{\!_C}_be^b)(u^a+e^a)$, with $e^a$ the Newtonian
label. Carry the CDM-threaded temperature over to that label by composing the
two,
\begin{equation}
  \hat\tau(x,e) \equiv \tau_{\!_C}\bigl(x,e_{\!_C}(e)\bigr)
  = \tau_{\!_N}(x,e) - v^{\!_C}_ae^a .
  \label{eq:slot}
\end{equation}
The aberration has gone. It was a change of argument and the change of argument
has now been made; what is left is the Doppler shift alone, which is a dipole.
Equation~(\ref{eq:slot}) is that composition at the order kept. The exact
statement is multiplicative: with $\mathcal{D}=\gamma(1-v^{\!_C}_ae^a)$ the
Doppler factor of Eq.~(\ref{eq:doppler}), one distribution function is split by
the two threadings as $\Pi_{\!_C}=\mathcal{D}^4\Pi_{\!_N}$, so the directional
temperatures are related by $\mathcal{D}$ and $1+\tau_{\!_C}$ by
$\mathcal{D}(1+\tau_{\!_N})$, up to the normalisation that fixes the directional
mean. Dropping from that the modulation $-(v^{\!_C}_ae^a)\tau_{\!_N}$, which
carries no factor of the rank and is $O(\ep^2\ell^0)$, returns
Eq.~(\ref{eq:slot}).
In this common label the two threadings carry the same multipoles above
$\ell=1$,
\begin{equation}
  \hat\tau_{A_\ell} = \tau^{\!_N}_{A_\ell}\ \ (\ell\ge2) ,
  \qquad
  \hat\tau_a = \tau^{\!_N}_a - v^{\!_C}_a ,
  \label{eq:slotmult}
\end{equation}
which is Eq.~(37) of Ref.~\cite{GebbieDunsbyEllis1999} read as a statement
about the label rather than about the sky. Their sources therefore differ by
the rate of a dipole and by nothing that carries a factor of the rank: the
whole $O(\ep^2\ell)$ difference between Eqs.~(\ref{eq:newt-pair})
and~(\ref{eq:cdm-triple}) is the difference between the two direction labels.

\emph{The relabelling.} The CDM hierarchy is written in the CDM label, so what
it propagates is not $\hat\tau_{A_\ell}$ but the multipoles of the same
function taken against $O^{A_\ell}(e_{\!_C})$. Undoing Eq.~(\ref{eq:slot}),
$\tau_{\!_C}(x,e)=\hat\tau(x,e)
+v^a_{\!_C\perp}\,\partial\hat\tau/\partial e^a$ to the order kept, so the two
sets of multipoles differ by $v^a_{\!_C\perp}\,\partial\tau/\partial e^a$. That
object needs no new reduction: it is Eq.~(\ref{eq:abersplit}) with
$K^a\to v^a_{\!_C}$, a vector carrying no direction dependence of its own,
which is the case already worked through in Sec.~\ref{sec:newtonian-pair}.
Hence, for $\ell\ge2$,
\begin{align}
  \tau^{\!_C}_{A_\ell} - \tau^{\!_N}_{A_\ell}
  &= -(\ell-1)\,v^{\!_C}_{\la a_\ell}\tau_{A_{\ell-1}\ra}
  \nonumber\\
  &\quad + \frac{(\ell+1)(\ell+2)}{2\ell+3}\,v_{\!_C}^b\,\tau_{bA_\ell} ,
  \label{eq:relabel}
\end{align}
with the Doppler dipole of Eq.~(\ref{eq:slotmult}) added to it at $\ell=1$. The
$\tau_{A_\ell}$ on the right may be read in either threading: the two differ by
the left-hand side, and the difference of the differences is of higher order.

The right-hand side is Eq.~(\ref{eq:newt-pair}) with $A^a\to v^a_{\!_C}$. The
relabelling and the Newtonian $O(\ep^2\ell)$ source are the same operator ---
one a finite displacement of the direction label, the other its rate --- and by
Eq.~(\ref{eq:twoframes}) both are built from the one boost: the source from its
rate of change $\dot v^{\!_C}_a+Hv^{\!_C}_a$, the relabelling from the boost
itself. That is how two threadings can differ in their sources at
$O(\ep^2\ell)$ without differing in their skies.

\emph{The consequence.} Equation~(\ref{eq:relabel}) is algebraic and pointwise:
no integral along the ray enters it and no derivative appears in it. What the
two threadings do at the ends of the ray is not ours to arrange --- each
congruence is defined by the matter it follows, so the relative velocity
$v^a_{\!_C}$ at emission and at reception is physical, and is not zero.

What makes the agreement exact is that $f$ is a scalar on the photon's phase
space. Both threadings split one photon at a shared event, and
$\Pi=\int E^3f\,\dd E$, so three powers of $\mathcal{D}$ come from $E^3$ and one
from the measure: their brightnesses there are related by
$\Pi_{\!_C}=\mathcal{D}^4\Pi_{\!_N}$ with nothing dropped --- the multiplicative relation above Eq.~(\ref{eq:slot}), not
Eq.~(\ref{eq:relabel}), which is truncated with everything else. The threading
dependence of the sky is therefore an endpoint Doppler and aberration
transformation, carried by $v^a_{\!_C}$ at the two ends and by nothing between
them, and Eq.~(\ref{eq:relabel}) is what it leaves on the multipoles at
$O(\ep^2\ell)$. Those endpoint transformations are not a residual dependence on
the intermediate threading, but they need not be $O(1)$ in multipole space. At
reception $v^a_{\!_C}$ is the observer's velocity relative to the threading
\cite{Planck2013XXVII}, and Eq.~(\ref{eq:relabel}) gives the standard
$O(v_{\!_R}\ell\tau)$ aberration mixing between neighbouring high-$\ell$
multipoles --- the rank growth that Table~\ref{tab:physics} assigns to observer
aberration. It leaves the full-sky $C_\ell$ unchanged at first order in a fixed
observer velocity, but the untransformed multipoles in two endpoint frames are
not frame-independent. At emission it is the baryon--CDM relative velocity,
acting on a field carrying $\ell\le2$ in the baryon frame, so that endpoint
coupling is $O(1)$ --- the count of Sec.~\ref{sec:residue}. Once both
predictions are carried to the same fixed physical emitter and observer frames
the observed sky is unique: what is independent of the intermediate threading is
the physical sky in fixed endpoint frames, not each threading's raw endpoint
multipoles.

\emph{Why not the chain rule.} The same content can be put differentially, as
the rate at which Eq.~(\ref{eq:relabel}) changes along the ray, and
Sec.~\ref{sec:cdm-triple} states it in that form. That route is the longer one.
Differentiating the relabelling at a fixed basis gives
\begin{equation}
  \frac{\dd}{\dd t}\bigl(v^a_\perp\bigr)
  = \Bigl(\frac{\dd v^a}{\dd t}\Bigr)_{\!\perp} ,
  \label{eq:dvperp}
\end{equation}
exactly: in the background $\dd(p^a/E)/\dd t=H(u^a+e^a)$ and
$\dd u^a/\dd t=He^a$, so $\dd e^a/\dd t=Hu^a$, and the screen projector
annihilates $u^a$ and $e^a$ alike. The answer is then short of
Eq.~(\ref{eq:dK}) by $Hv^a_\perp$. That term is not in the boost; it is in the
screens. Each threading carries its own, and the two differ at first order by
$v^au_b+u^av_b$, so transporting the photon's direction in one and reading it in
the other leaves a term in $v^a_\perp$ whose coefficient is the background
redshift rate, $-EH$ by Eq.~(\ref{eq:redshiftcoef}). That is the third term of
App.~\ref{app:shift} and it is the whole of the $Hv$: one direction index, with
weight $H$, and not one contribution per multipole index. Differentiating the
relabelling at a fixed basis is
the same error there as differentiating Eq.~(\ref{eq:aber}) at a fixed
projector. Equation~(\ref{eq:relabel}) carries none of that bookkeeping, which
is why it is the statement given here.

\subsection{The aligned limit}

\label{sec:aligned}

The redistributive character of the coupling is already visible in a single
mode. Take the acceleration or shear mode and the temperature mode on a common
axis $\hat k$, and write $\tau(e)=\sum_\ell c_\ell P_\ell(\mu)$,
$\mu=\hat k\cdot e$. With $A_a=A\hat k_a$ and
$\sigma_{ab}=\sigma\,\hat k_{\la a}\hat k_{b\ra}$ the aberration operators of
Eq.~(\ref{eq:aberration}) are $A(1-\mu^2)\partial_\mu$ and
$\sigma\mu(1-\mu^2)\partial_\mu$, and at leading order in $1/\ell$
\begin{align}
  \dot c_\ell\big|_{\mathrm{N},A}
    &\simeq \ts\frac{\ell}{2}\,A\,\bigl(c_{\ell+1}-c_{\ell-1}\bigr) ,
    \nonumber\\
  \dot c_\ell\big|_{\mathrm{C},\sigma}
    &\simeq \ts\frac{\ell}{4}\,\sigma\,\bigl(c_{\ell+2}-c_{\ell-2}\bigr) ,
    \qquad \ell\gg1 .
  \label{eq:legendre}
\end{align}
The label is the threading each rate is computed in. They are alternative
representations of the same transport, not two contributions to one rate, and
both are finite differences in $\ell$.

That route passes through a Legendre expansion, and the whole point of the
claim is the relative sign, so it is better had without one. The aligned
configuration is specified covariantly: $A_a=A\hat k_a$,
$\sigma_{ab}=\sigma\hat k_{\la a}\hat k_{b\ra}$ and
$\tau_{A_\ell}=\tau_\ell\,\hat k_{\la A_\ell\ra}$, with $\hat k^a\hat k_a=1$.
Write $\Lambda_\ell\equiv\hat k_{\la A_\ell\ra}O^{A_\ell}$, a scalar on the
sphere, and note that two identities already in hand do everything. First,
C6.1 read on $\hat k$ gives
$\hat k^b\hat k_{\la bA_{\ell-1}\ra}=\frac{\ell}{2\ell-1}\hat k_{\la A_{\ell-1}\ra}$;
and because $\hat k_{\la A_\ell\ra}$ is trace-free, C5.8 gives
$e^b\hat k_{\la bA_{\ell-1}\ra}O^{A_{\ell-1}}=\Lambda_\ell$. Second,
Eq.~(\ref{eq:red-A}) read on $\hat k$ is the recursion
\begin{equation}
  \mu\,\Lambda_m = \Lambda_{m+1} + r_m\Lambda_{m-1} ,
  \qquad r_m\equiv\frac{m^2}{4m^2-1} ,
  \label{eq:lamrec}
\end{equation}
with $\mu\equiv\hat k_ae^a$; at $m=1$ it returns $\Lambda_2=\mu^2-\frac13$, as
the definition requires.

In the Newtonian threading $K^a=A\hat k^a$ and Eq.~(\ref{eq:abersplit}) gives,
at rank $\ell$,
\begin{equation}
  \frac{(\ell+1)^2(\ell+2)}{(2\ell+1)(2\ell+3)}\,\tau_{\ell+1}
  - (\ell-1)\,\tau_{\ell-1} .
  \label{eq:alignN}
\end{equation}
In the CDM threading $K^a=\sigma\hat k^{\la a}\hat k^{b\ra}e_b$, so
$K_be^b=\sigma\Lambda_2$ and the two $\frac13$'s cancel between the transverse
and longitudinal sums, leaving $\sum_n n\tau_n[\frac{n}{2n-1}\mu\Lambda_{n-1}
-\mu^2\Lambda_n]$; applying Eq.~(\ref{eq:lamrec}) twice gives, at rank $\ell$,
\begin{equation}
  \frac{(\ell+1)^2(\ell+2)^2(\ell+3)}{(2\ell+1)(2\ell+3)^2(2\ell+5)}\,\tau_{\ell+2}
  - (\ell-2)\,\tau_{\ell-2} ,
  \label{eq:alignC}
\end{equation}
together with a rank-preserving term whose coefficient is
$\ell[\frac{\ell}{2\ell-1}-r_{\ell+1}-r_\ell]$ and therefore $O(1)$. These are
exact, and their high-$\ell$ limits are the $\frac14$ and $\frac1{16}$ of
Eq.~(\ref{eq:modeform}). The $\frac12$ and $\frac14$ of
Eqs.~(\ref{eq:legendre}) are those same numbers after the $\beta_\ell$
conversion below.

The cancellation then needs no expansion either. Multiply
Eqs.~(\ref{eq:alignN}) and~(\ref{eq:alignC}) by $\beta_\ell$ and use C5.3 to
move the $\beta$'s onto the multipoles. Both become finite differences in the
one variable $\beta_\ell\tau_\ell$:
\begin{equation}
\begin{gathered}
  \frac{(\ell+1)(\ell+2)}{2\ell+3}\,\beta_{\ell+1}\tau_{\ell+1}
  - \frac{\ell(\ell-1)}{2\ell-1}\,\beta_{\ell-1}\tau_{\ell-1} ,
  \\
  \frac{(\ell+1)(\ell+2)(\ell+3)}{(2\ell+3)(2\ell+5)}\,\beta_{\ell+2}\tau_{\ell+2}
  \\
  \qquad\qquad
  - \frac{\ell(\ell-1)(\ell-2)}{(2\ell-1)(2\ell-3)}\,\beta_{\ell-2}\tau_{\ell-2} ,
\end{gathered}
  \label{eq:aligndiff}
\end{equation}
Write $f_\ell\equiv\beta_\ell\tau_\ell$, with $\beta_\ell$ of
[Table~\ref{tab:conventions}, C5.2], and split each bracket into the part that
is a difference and the part that is not,
\begin{align}
  c_+f_{\ell+n} - c_-f_{\ell-n}
  &= \frac{c_++c_-}{2}\bigl(f_{\ell+n}-f_{\ell-n}\bigr)
  \nonumber\\
  &\quad + \bigl(c_+-c_-\bigr)f_\ell + \ldots
  \label{eq:stencil}
\end{align}
The first part is a derivative stencil: $n=1$ at weight
$\frac12(c_++c_-)\to\ell/2$ for the Newtonian bracket, $n=2$ at weight
$\to\ell/4$ for the CDM one. In both the width times the weight is one,
$2\times\frac12=4\times\frac14=1$, so both stencils are unit-normalised and
both reconstruct the same quantity, $\ell\,\partial_\ell f_\ell$, for $f_\ell$
slowly varying across the wider stencil, $\Delta\ell=4$. That condition is on
the sequence and not on the coefficients, whose closure below is exact. The two
threadings do not merely both cancel: each is the same derivative, sampled on a
different stencil. That is Sec.~\ref{sec:bulkendpoint}'s statement made
quantitative --- a remapping acts on the multipole index as a dilation, so it
returns a derivative in $\ell$ and not a multiple of the amplitude.

The second part of Eq.~(\ref{eq:stencil}) does not vanish, and it is where the
two threadings would part company if anything did:
\begin{equation}
  c_+-c_- =
  \frac{2(2\ell^2+2\ell-1)}{(2\ell-1)(2\ell+3)} \;\to\; 1
  \label{eq:resN}
\end{equation}
for the Newtonian bracket, against
\begin{equation}
  c_+-c_- =
  \frac{3\,(4\ell^4+8\ell^3-11\ell^2-15\ell+6)}
       {(2\ell-3)(2\ell-1)(2\ell+3)(2\ell+5)}
  \;\to\; \tfrac34
  \label{eq:resC}
\end{equation}
for the CDM one. The missing quarter is the rank-preserving coupling. It exists
in the CDM threading and not in the Newtonian one, for a structural reason:
$A^a$ carries no direction dependence of its own and so couples $\ell$ to
$\ell\pm1$ and to nothing else, while $\sigma^a{}_be^b$ does and so reaches
$\ell$ itself. Its coefficient, from the $n=\ell$ term above, is
\begin{equation}
  \ell\Bigl[\frac{\ell}{2\ell-1}-r_{\ell+1}-r_\ell\Bigr]
   = \frac{\ell(\ell+1)}{(2\ell-1)(2\ell+3)} \;\to\; \tfrac14 ,
  \label{eq:rankpres}
\end{equation}
and $\frac34+\frac14=1$. The two brackets therefore agree at $O(1)$ as well as
at $O(\ell)$, by two coefficients of the one reduction rather than by two
independent routes: the stencil identity gives the first order, and
Eq.~(\ref{eq:rankpres}) --- the aligned-limit coefficient of the
rank-preserving coupling, which only the CDM threading has --- gives the
second. What is left is
$O(\ell^{-2})$ --- the brackets carry different curvature content, as stencils
of width two and width four must --- so the agreement is complete at both
orders the decomposition resolves. What this establishes is the angular
algebra: that the two divisions carry the same operator, which is what
Sec.~\ref{sec:relabel} needs. That the two threadings give the same sky is that
section's exact boost, not this. Figure~\ref{fig:threadings}(b) shows the closure
and what is left of it.

In the aligned configuration the brackets of
Eq.~(\ref{eq:modeform}) are $\ell[-\tau_{\ell-1}+\frac14\tau_{\ell+1}]$ in the
Newtonian threading and $\ell[-\tau_{\ell-2}+\frac1{16}\tau_{\ell+2}]$ in the
CDM threading. With $\tau_{A_\ell}=c_\ell n_\ell\hat k_{\la A_\ell\ra}$ and
$n_\ell=(2\ell-1)!!/\ell!=\beta_\ell^{-1}$, the ratios
$n_{\ell\mp1}/n_\ell\to\frac12,2$ and $n_{\ell\mp2}/n_\ell\to\frac14,4$ convert
the two brackets into Eqs.~(\ref{eq:legendre}). The same result follows
directly from the Legendre identity
$(1-\mu^2)P'_\ell = \frac{\ell(\ell+1)}{2\ell+1}(P_{\ell-1}-P_{\ell+1})$.

The cancellation in the Newtonian threading was first obtained in
Ref.~\cite{GebbieThesis1999}, by the route that follows. It is recorded because
it shows where the $\beta_\ell$ ratios enter, not because anything above rests
on it. Inserting the almost-FLRW integral
solutions and applying the weak-coupling approximation \cite{HuWhite1996}, the
two terms of the bracket weight $\Phi_A(k,\eta_{\ell-1})$ and
$\Phi_A(k,\eta_{\ell+1})$ by
\begin{equation}
\begin{gathered}
  \frac{(2\ell-1)\beta_\ell}{\beta_{\ell-1}\sqrt{\ell-1}}
   = \frac{\ell}{\sqrt{\ell-1}} \to \sqrt\ell ,
  \\
  \frac14\,\frac{(2\ell+3)\beta_\ell}{\beta_{\ell+1}\sqrt{\ell+1}}
   = \frac{(2\ell+3)(2\ell+1)}{4(\ell+1)^{3/2}} \to \sqrt\ell ,
\end{gathered}
\end{equation}
equal and opposite [Table~\ref{tab:conventions}, C5.3; the section term is added
in App.~\ref{app:endpoints}], so the two cancel
when $\eta_{\ell-1}\simeq\eta_{\ell+1}$, which holds at high $\ell$. The
cancellation is exact at leading order in $1/\ell$ and independent of the
damping factor: the $\frac14$ of the hierarchy is what cancels the $\beta_\ell$
ratio of $4$. The CDM-frame bracket cancels in the same way, the two-step ratios
compensating the $\frac1{16}$, because the $\beta_\ell$ ratios are exactly what
converts PSTF to Legendre coefficients.

\section{The lensing contribution to $C_\ell$ is quartic}
\label{sec:quartic}

The aberration source is a product of two first-order fields:
$K_\perp^a\propto\D\Phi$ and $\partial_e\tau^{(1)}\propto\Phi$. So
$\delta\tau^{\rm NL}\propto\Phi\Phi$, and its contribution to $C_\ell$ is
\emph{quartic} in the potential, where the linear spectrum is quadratic. That
fact is necessary for the $O(\ell)$ growth of the source not to reach the power
as $\ell^2$; it is not sufficient. The rest is a cancellation between the two
fourth-order terms.

On scales small compared with the sky, Eq.~(\ref{eq:deltax}) makes the observed
anisotropy a remapping of the emitted one,
$\tau_R(\bm\theta)=\tau(\bm\theta+\bm\alpha)$, with deflection angle
$\bm\alpha=\delta x_\perp/\chi_*=\nabla_{\!\theta}\psi$ and
\begin{equation}
  \psi(\bm\theta) = -\int_0^{\chi_*}\!\dd\chi\,
  \frac{\chi_*-\chi}{\chi_*\,\chi}\,(\Phi_A-\Phi_H)\bigl(\chi\bm\theta,\eta_0-\chi\bigr) ,
  \label{eq:psi}
\end{equation}
using $\nabla_{\!\perp}=\chi^{-1}\nabla_{\!\theta}$ at distance $\chi$. To the
order needed,
\begin{equation}
  \tau_R = \tau^{(1)}
  + \underbrace{\alpha^a\nabla_{\!a}\tau^{(1)}}_{\delta\tau^{\rm NL}}
  + \underbrace{\ts\frac12\alpha^a\alpha^b\nabla_{\!a}\nabla_{\!b}\tau^{(1)}}_{\tau^{(3)}}
  + O(\ep^4) .
  \label{eq:remap}
\end{equation}
The middle term is the line-of-sight integral of Sec.~\ref{sec:lensing}; the
last is $O(\ep^3)$ and is not formed by a calculation carried to $O(\ep^2)$.

\emph{The cross term vanishes, under a premise.}
$\langle\tau^{(1)}\delta\tau^{\rm NL}\rangle$ is a three-point function. The
problem is statistical, not algebraic. Here we assume that the deflection is
statistically \emph{independent} of the primary anisotropy, which is what
factorising a three-point expectation needs and is strictly more than the
vanishing correlation used elsewhere, giving
$\langle\bm\alpha\rangle\cdot\langle\tau\nabla\tau\rangle=0$. Joint Gaussianity
would do it outright, every three-point function of jointly Gaussian fields
being zero. Neither premise implies the other: independent fields need not be
Gaussian, and jointly Gaussian fields may be correlated. The leading
correlation of the late-time potentials with the integrated Sachs--Wolfe
contribution is a bispectrum, and does not enter $C_\ell$ at this order. So
$\delta\tau^{\rm NL}$, although $O(\ep^2)$ in the field, changes $C_\ell$ first
at $O(\ep^4)$.

\emph{The fourth-order terms.} Writing $C^\psi_L$ for the spectrum of $\psi$ and
\begin{equation}
  R \equiv \ts\frac12\bigl\langle|\bm\alpha|^2\bigr\rangle
  = \frac{1}{4\pi}\int\!\dd L\,L^3C^\psi_L ,
\end{equation}
there are two contributions \cite{LewisChallinor2006},
\begin{equation}
\begin{split}
  \Delta C_\ell
  &= \bigl\langle|\delta\tau^{\rm NL}|^2\bigr\rangle_\ell
   + 2\bigl\langle\tau^{(1)}\tau^{(3)}\bigr\rangle_\ell \\
  &= \int\!\frac{\dd^2\ell'}{(2\pi)^2}\,
     \bigl[\bm\ell'\cdot(\bm\ell-\bm\ell')\bigr]^2
     C^\psi_{|\bm\ell-\bm\ell'|}\,C_{\ell'} \\
  &\quad - \ell^2R\,C_\ell .
\end{split}
  \label{eq:dCl}
\end{equation}
Each is $O(\Phi^4)$. Under $\Phi\to\lambda\Phi$, $\Delta C_\ell\to\lambda^4\Delta C_\ell$
while $C_\ell\to\lambda^2C_\ell$, so no relation of the form
$\Delta C_\ell\sim\ell^n\times(\text{a spectrum quadratic in }\Phi)$ can hold at
more than one amplitude.

\emph{The $\ell^2$ cancels.} The first term of Eq.~(\ref{eq:dCl}), taken alone,
grows as $\ell^2$: for lens multipoles $L=|\bm\ell-\bm\ell'|\ll\ell$ it tends to
$+\ell^2R\,C_\ell$. That is the square of the explicit $\ell$ in
Eq.~(\ref{eq:nl-source}). The second term is exactly $-\ell^2R\,C_\ell$, and the
two cancel. Physically, a uniform deflection translates the sky patch rigidly
and cannot change its power spectrum. What remains is weighted by the gradient
of $\bm\alpha$, the convergence and image distortion of the lens map, and is
suppressed by $(L/\ell)^2$ relative to the cancelled terms. At leading order in
$L/\ell$ the remainder is
\begin{equation}
  \Delta C_\ell \simeq \frac{1}{2\pi}\int\!\dd L\,L^5C^\psi_L\,
  \Bigl[\ts\frac{3}{16}\ell^2C_\ell'' + \ts\frac{17}{16}\ell C_\ell' + C_\ell\Bigr] ,
  \label{eq:remainder}
\end{equation}
which depends on the logarithmic slope and curvature of $C_\ell$; the three
coefficients are derived in App.~\ref{app:quartic-remainder}. The expansion
that produces it is in $L/\ell$, so Eq.~(\ref{eq:remainder}) is the residual
where the deflection power sits at $L\ll\ell$ and $C_\ell$ varies slowly across
the displacement; where it does not --- at a sharp acoustic feature, or in the
damping tail fed by power from much larger scales --- the two $\ell^2$ terms do
not cancel multipole by multipole, and what survives is the redistribution
below. For a power law
$C_\ell\propto\ell^{-n}$ the bracket is $(n-2)(3n-8)C_\ell/16$, so a
scale-invariant spectrum, $\ell^2C_\ell=\text{const}$, is unchanged at this
order.

\emph{Power is conserved.} For a statistically homogeneous primary field
independent of the deflection, evaluation at the remapped direction preserves
the ensemble one-point variance, $\langle\tau_R^2\rangle=\langle\tau^2\rangle$,
so the integrated fourth-order correction vanishes:
$\int\dd^2\ell\,\Delta C_\ell=0$, which Eq.~(\ref{eq:dCl}) satisfies
identically. Lensing redistributes power between multipoles --- it smooths the
acoustic peaks and fills the damping tail --- and adds no ensemble variance. In
that sense the remapping only moves values around the sky.

The variance statement does not require the lens map to be globally one-to-one.
For each observed direction $\bm\theta$ ray tracing still selects one source
direction $\bm\beta(\bm\theta)$ and preserves surface brightness; a caustic
occurs where the Jacobian
$\mathcal{A}^i{}_j=\partial\beta^i/\partial\theta^j$ becomes singular,
$\det\mathcal{A}=0$, so that the inverse map from a source direction to its
images is multivalued. No new local term is then added to the Liouville
operator, and no temperatures from distinct images are summed at one observed
direction. What fails is the globally invertible weak-deformation description,
and the operator identification survives caustic formation. The accuracy of the
finite expansion (\ref{eq:remap}) is a separate question, and $\det\mathcal{A}$
does not control it: a coherent deflection has vanishing deformation gradient
and is a rigid rotation of the sky, which leaves $C_\ell$ unchanged. What
governs Eqs.~(\ref{eq:dCl})--(\ref{eq:remainder}) is the deflection relative to
the separation being resolved --- the condition already imposed above, that the
spectrum be smooth on the deflection scale.

This is the structural content of the $O(\ell)$ couplings for the power
spectrum. The source grows as $\ell$, as the exact hierarchy says. The power
does not grow as $\ell^2$, because the growth enters at fourth order in the
field, in two terms of opposite sign, and only its gradient survives.

Both halves of that are needed. The nonlinear source \emph{matters}: it is where
an effect normally supplied from outside the hierarchy is derived inside it, in
the variables the hierarchy is already written in; and it adds nothing, its
contribution being quartic in the potential and equal to standard lensing.
Stating the first alone invites a new effect; stating the second alone makes the
derivation look like an exercise.


\section{Discussion}
\label{sec:discussion}

The identification of these couplings is not new. The second-order Boltzmann
hierarchy of Ref.~\cite{BenekeFidler2010} contains them, with coefficients
growing like $\ell$, and names them as weak lensing; the $\ell$-growth is
remarked on there. Reference~\cite{Pettinari2013} and the code built on it keep
the couplings through recombination and then exclude them from the late-time
line-of-sight sources, deferring to the standard remapping.
Reference~\cite{Pitrou2009} comes closer. It gives the second-order Boltzmann
equation with polarization in a tetrad formalism, expands it in the same PSTF
basis $n^{\la a_\ell\ra}$ used here [its Eq.~(8.1)], carries the same aberration
operator in its second-order Liouville operator, where it is named as lensing,
and writes the transformation of the radiation multipoles under a change of
observer velocity exactly, as an angular integral [its Eq.~(2.49)]. The
hierarchy itself is computed in the Newtonian gauge [its Eq.~(5.19)], so that
transformation relates tetrads at a point rather than threadings inside the
equations; the frame is fixed before the hierarchy is written. It also cites
Ref.~\cite{MaartensGebbieEllis1999} and corrects a coefficient in its Eq.~(60),
which is not among the equations used here. None of the three
works in the $1+3$ covariant variables; Ref.~\cite{Pitrou2009} names that an
alternative programme and does not follow it.

All three formulate the hierarchy in a scalar shear-free representation and fix
the observer tetrad before solving it. In the terminology used here that is the
Newtonian-threading representation: the scalar-mode congruence is shear-free,
every $O(\ell)$ coupling is an $\ell\pm1$ gradient coupling, and there is no
$\ell\pm2$ shear coupling beside it. With the frame fixed, the frame dependence of the couplings cannot be
stated, and the observer's aberration --- the frame-dependent part of the same
operator --- does not arise.

What is added here is the frame. In an arbitrary threading the $O(\ell)$
couplings are one operator, $K_\perp^a\,\partial/\partial e^a$
[Eq.~(\ref{eq:aberration})]. Under a change of threading it shifts by the
generator of the event-by-event aberration map [Eq.~(\ref{eq:dK})], and the
shift is undone on the multipoles by a pointwise algebraic relabelling
[Eq.~(\ref{eq:relabel})], so that the observed sky is unchanged at every $\ell$
and not only at the order at which the couplings are kept. Its content
divides into lensing, which is threading-independent because it is sourced by
$E_{ab}$, and observer aberration, which is fixed by $u_R^a$. The explicit
coupling is carried by the acceleration in the Newtonian threading and by the
shear in the CDM threading. We are not aware of a treatment of these couplings that
states their frame covariance, or that places observer aberration in the same
operator as lensing. The $1+3$ variables are what make the statement possible:
the congruence is fixed inside the equations and can therefore be changed inside
them, with no second-order gauge construction.

Gauge-invariant metric perturbation theory
is usually described as an approximate representation of general relativity, and
within its domain it is one. The reduction is more constraining than the
description usually suggests. Once a background metric, a temporal foliation and an identification
between the physical and the background spacetime are fixed, the resulting
system is no longer free to compare arbitrary observer congruences within
itself. Effects that a change of physical threading redistributes between
acceleration, shear, spatial curvature, Doppler terms and angular aberration are
not displayed as an internal covariance of that reduced hierarchy; they must be
recovered by an explicit observer or tetrad transformation, or by deriving the
corresponding higher-order system. Its
variables are gauge invariant relative to the chosen background map, and the
formulation is not frame covariant under a change of the physical observer
congruence; in that precise sense it is a derived system with a smaller
covariance structure than the theory it approximates. What is lost is not
coordinate dependence, which was the thing being removed, but the freedom to
express the same radiation transport relative to different physical threadings
and to verify their equivalence inside the formulation. The $O(\ell)$ couplings
are a case in point. Removed by linearising the free-streaming term, they cannot
be recovered from the linear hierarchy or from its solution; they are restored
either by carrying the hierarchy to the next order or by a geodesic remapping
imposed from outside it, and in practice it is the second
\cite{LewisChallinor2006,Pettinari2013}.

The $1+3$ route keeps that freedom in place. The radiation field, the photon
direction and the kinematics are defined relative to a physical congruence in
the exact Einstein--Liouville system, and the approximation is made after that;
the almost-Friedmann--Lema\^itre ordering is an observer-anchored statement
about measured near-isotropy rather than a prior restriction on how the
spacetime is represented. Acceleration, shear, vorticity, spatial curvature and
direction-space transport stay inside one system, and a change of threading acts
as a Doppler--aberration relabelling of the same observable sky,
Eq.~(\ref{eq:relabel}). None of this makes the metric treatment wrong where it
applies. It says why effects excluded when the metric and the time surfaces are
fixed cannot always be recovered from the reduced hierarchy alone, and what is
kept by not fixing them first.

The route is a statement about derivation, not a recommendation for computation.
It places lensing inside the hierarchy that is supposed to contain it, without
second-order gauge-invariant variables. It is not a proposal to obtain lensing
by integrating a truncated hierarchy: the couplings connect $\ell$ to $\ell\pm1$
and $\ell\pm2$ with coefficients growing in $\ell$, so a truncation converges
slowly. That is the practical reason lensing is computed as a remapping
\cite{LewisChallinor2006}, and the reason second-order codes exclude these terms
from the late-time sources \cite{Pettinari2013}. How the truncated hierarchy
converges, and where it fails to, is a question this formulation poses and does
not answer here.

Two further things are posed and not settled. The frame statement is established
here by derivation, and a derivation checked against a derivation is one source;
a solver carrying the threading as a parameter, integrating the same
configuration along two of them and comparing the observables, would be a
second. No existing code is that: in each, the gauge choice sets the congruence
the equations are written relative to, so the threading is not a free parameter.
That is why none of the treatments above carries an $\ell\pm2$ shear coupling. And the tensor contribution is placed here but not evaluated:
its operator and its $\ell$-weight are the ones already in
Table~\ref{tab:coefficients}, while its amplitude needs the tensor sector's own
evolution, and the polarization it would be observed through is not carried by
this formulation [App.~\ref{app:tensor}].

The exact $1+3$ hierarchy retains an observer-threading freedom that a fixed
metric perturbative hierarchy does not display internally, and the $O(\ell)$
couplings are the frame-covariant angular-transport operator through which
lensing and the observer's own aberration appear in it.

\begin{acknowledgments}
I thank George Ellis, Roy Maartens and the late Fr.~Bill Stoeger SJ for many
conversations over the years.
\end{acknowledgments}

\section*{Disclosure of AI assistance}

A large language model (Claude, Anthropic) was used in preparing this work --- in checking
the derivations, in the symbolic checking and in the drafting based on the initial drafts and calculations. What it did, and
where, is recorded section by section in the reproducibility bundle, together
with the audits run against it.

All physical arguments, derivations and conclusions have been checked by the
author, who takes full responsibility for them. The model is not an author and
does not satisfy authorship criteria. No text was published without author
review.


\section*{Reproducibility}

The reproducibility bundle that accompanies this paper, version
\texttt{v1.0.0} \cite{Bundle}, carries the conventions sheet on which every sign
and $\ell$-weight in the paper depends, the symbolic checks of the contraction
identities and coupling coefficients, the scripts that produce each figure, and
the supplementary material.

\bibliographystyle{apsrev4-2}
\bibliography{TG-FrameCovariantCMBR-v1.1.6-arxiv}

\appendix

\section{Conventions}
\label{app:conventions}

Table~\ref{tab:conventions} is the sheet every sign and $\ell$-weight in the
body cites.

\begin{table*}[tp]
\caption{Conventions. Every sign and every $\ell$-weight in this paper rests on
one of these lines. A \emph{definition} is adopted from the source named;
a \emph{result} is derived from the definitions and checked under them; a
\emph{stipulation} is a choice no source consulted here fixes. The supplementary material
\cite{Bundle} restates the couplings in the conventions of the other
treatments, and carries the correspondence between these numbers and the
preprints'. Equation numbers are those of the \emph{published} papers.}
\label{tab:conventions}
\begin{ruledtabular}
\begin{tabular}{l>{\raggedright\arraybackslash}p{0.49\textwidth}>{\raggedright\arraybackslash}p{0.26\textwidth}l}
 & Statement & Source & Kind \\
\colrule
C0.1 & $A=B$ exact; $A\simeq B$ almost-Friedmann--Lema\^itre; $A\approx B$ to a further smallness parameter
     & \cite{GebbieDunsbyEllis1999} \S3.2 and App.~G & definition \\
C1.1 & Signature $(-,+,+,+)$; $u^au_a=-1$, $e^ae_a=+1$
     & \cite{MaartensGebbieEllis1999}, \cite{GebbieThesis1999} & definition \\
C1.2 & $h_{ab}=g_{ab}+u_au_b$; positive semi-definite on the rest space
     & \cite{MaartensGebbieEllis1999} & definition \\
C1.5 & $p^a=E(u^a+e^a)$, $e^ae_a=+1$, $e^au_a=0$
     & \cite{MaartensGebbieEllis1999}, \cite{GebbieEllis1998} & definition \\
C1.6 & Contracting a pair of spatial indices introduces no sign, $h_{ab}$ being positive semi-definite on the rest space
     & C1.1 and C1.2 & result \\
C2.1 & $\D_c = h_c{}^d\nabla_d$, acting on an already projected tensor
     & \cite{MaartensGebbieEllis1999}, \cite{EllisVanElst1998} & definition \\
C2.2 & $\nabla_bu_a = -A_au_b + \frac13h_{ab}\Theta + \ep_{abc}\omega^c + \sigma_{ab}$
     & \cite{MaartensGebbieEllis1999}, \cite{GebbieThesis1999} & definition \\
C2.3 & $E_{ab}=C_{acbd}u^cu^d$, $H_{ab}=\frac12\ep_{acd}C^{cd}{}_{be}u^e$
     & \cite{Ellis1971}, \cite{EllisVanElst1998} & definition \\
C2.4 & $R^a{}_{bcd}=\partial_c\Gamma^a{}_{db}-\partial_d\Gamma^a{}_{cb}+\Gamma^a{}_{ce}\Gamma^e{}_{db}-\Gamma^a{}_{de}\Gamma^e{}_{cb}$, $R_{bd}=R^a{}_{bad}$
     & the convention under which \cite{EllisVanElst1998} Eq.~(55) gives ${}^{(3)}\!R=6K/a^2$ for FLRW & result \\
C2.5 & $E_{ab}-\frac12\pi_{ab}\approx\D_{\la a}\D_{b\ra}\Phi_A$, $E_{ab}+\frac12\pi_{ab}\approx-\D_{\la a}\D_{b\ra}\Phi_H$
     & \cite{GebbieDunsbyEllis1999} Eq.~(26) & definition \\
C2.6 & Shear-free frame: $\D_{\la a}A_{b\ra}\approx E_{ab}-\frac12\pi_{ab}$, so $A_a\approx\D_a\Phi_A$
     & \cite{GebbieDunsbyEllis1999} Eq.~(49); \cite{EllisVanElst1998} Eq.~(31) at $\sigma_{ab}=0$ & result \\
C2.7 & ${}^{(3)}\!R_{\la ab\ra}=E_{ab}+\frac12\pi_{ab}$, shear-free and irrotational; exact, not linearised
     & \cite{EllisVanElst1998} Eqs.~(54) and~(31), Eq.~(\ref{eq:gauss-tf}) & result \\
C3.1 & $\D^a\D_aQ=-k^2Q$, $k$ physical, $Q$ time-independent
     & \cite{GebbieEllis1998} Eq.~(47) & definition \\
C3.2 & $Q_{A_\ell}=(-k)^{-\ell}\D_{\la A_\ell\ra}Q$; stripped factor real, no $i$
     & \cite{GebbieEllis1998} Eq.~(48) & definition \\
C3.4 & $\tau_{A_\ell}=\sum_k\tau_\ell(k)Q_{A_\ell}$, with $Q_{A_\ell}$ as in C3.2
     & \cite{GebbieEllis1998} Eq.~(60), on the harmonics of its Eq.~(48) & definition \\
C3.5 & $X=-\sum_kX(k)Q$, forced by Eq.~(\ref{eq:scalarmode}) under C3.2
     & no source expands a scalar & stipulated \\
C4.1 & $\D_aQ=-k\,Q_a$ & C3.2 at $\ell=1$ & result \\
C4.2 & $\D_{\la a}Q_{A_\ell\ra}=-k\,Q_{aA_\ell}$ (raising) & C3.2 & result \\
C4.3 & $\D^a\D_{\la a}Q_{A_\ell\ra}=\frac{\ell+1}{2\ell+1}(-k^2)Q_{A_\ell}$ at $K=0$
     & \cite{GebbieEllis1998} Eq.~(C.9) & result \\
C4.4 & $\D^aQ_{aA_\ell}=+\frac{\ell+1}{2\ell+1}k\,Q_{A_\ell}$ (lowering)
     & C3.2, \cite{GebbieEllis1998} Eq.~(137) & result \\
C4.5 & Mode recursion: free streaming couples $\ell\pm1$ with \emph{opposite} real signs, though both covariant couplings carry a plus
     & \cite{GebbieEllis1998} Eq.~(C.10) & result \\
C5.1 & $O^{A_\ell}=e^{\la A_\ell\ra}$, $\tau=\sum_\ell\tau_{A_\ell}O^{A_\ell}$
     & \cite{GebbieEllis1998} Eq.~(15) & definition \\
C5.2 & $\beta_\ell=O^{A_\ell}O_{A_\ell}=\ell!/(2\ell-1)!!$
     & \cite{GebbieEllis1998} Eq.~(23) & definition \\
C5.3 & $\beta_\ell/\beta_{\ell-1}=\ell/(2\ell-1)$, $\beta_\ell/\beta_{\ell+1}=(2\ell+1)/(\ell+1)$
     & \cite{GebbieEllis1998} Eq.~(24) & result \\
C5.5 & $\Delta_\ell=4\pi\beta_\ell/(2\ell+1)$; $\int\dd\Omega\,O^{A_\ell}O_{B_m}=\delta_{\ell m}\Delta_\ell h^{\la A_\ell\ra}{}_{\la B_\ell\ra}$
     & \cite{GebbieEllis1998} Eq.~(B.7) & definition \\
C5.7 & $h^{\la A_\ell\ra}{}_{\la A_\ell\ra}=2\ell+1$
     & \cite{GebbieEllis1998} Eq.~(B.9) & result \\
C5.8 & $\Pi_{A_\ell}O^{A_\ell}=\Pi_{A_\ell}e^{A_\ell}$ for PSTF $\Pi_{A_\ell}$
     & C5.1 and trace-freeness & result \\
C5.9 & $T^{\la B_m\ra}=T^{B_m}-\frac{m(m-1)}{2(2m-1)}h^{(b_1b_2}t^{B_{m-2})}$, $t$ the trace
     & App.~\ref{app:longer}, Eq.~(\ref{eq:detrace}) & result \\
C6.1 & $e_{a_1}O^{A_\ell}=\frac{\ell}{2\ell-1}O^{A_{\ell-1}}$, equivalently $e_aO^{aA_\ell}=\frac{\ell+1}{2\ell+1}O^{A_\ell}$
     & \cite{GebbieEllis1998} Eq.~(20) & result \\
C6.2 & $e_ae_bO^{abA_\ell}=\frac{(\ell+1)(\ell+2)}{(2\ell+1)(2\ell+3)}O^{A_\ell}$
     & C6.1 twice & result \\
C6.3 & $e^aO^{A_\ell}=O^{aA_\ell}+\frac{\ell}{2\ell+1}h^{a\la a_\ell}O^{A_{\ell-1}\ra}$ (raising)
     & C6.1, Eq.~(\ref{eq:eO}) & result \\
C6.7 & The $(\ell+1)/(2\ell+1)$ of C4.4 and that of C6.1 are different objects: a divergence of a gradient in space, and a contraction of $e$ with $O$ on the sphere
     & C4.4 and C6.1 & result \\
\end{tabular}
\end{ruledtabular}
\end{table*}

\section{Longer PSTF manipulations}
\label{app:longer}

The coefficient of Eq.~(\ref{eq:detrace}) is a counting argument. Write
$U^{B_m}=h^{(b_1b_2}t^{B_{m-2})}$ as the average over the $\binom{m}{2}$
distinct ways of choosing which index pair carries $h$. Tracing $U$ on $b_1b_2$,
the pair $\{b_1,b_2\}$ itself contributes $h_{b_1b_2}h^{b_1b_2}=3$; each of the
$m-2$ pairs $\{b_1,b_j\}$ with $j\ge3$ contributes $1$, as does each of the
$m-2$ pairs $\{b_2,b_j\}$; and every pair internal to $b_3\ldots b_m$ leaves a
trace of $t$, which vanishes. The surviving count is $3+2(m-2)=2m-1$, so
$h_{b_1b_2}U^{B_m}=\frac{2(2m-1)}{m(m-1)}t^{B_{m-2}}$ and the subtraction that
makes $T$ trace-free carries the reciprocal. The same count, with $h^{cd}$ in
place of one index and $\sigma_{cd}$ or $\Pi_{A_n}$ absorbing the rest, gives
the surviving pairings quoted below Eq.~(\ref{eq:vO}).

\section{The quartic remainder coefficients}
\label{app:quartic-remainder}

Mode expansions here are taken with the measure
$\sum_k\to V\!\int\!\dd^3k/(2\pi)^3$, and for an aligned isotropic convolution
$4\pi/(2\pi)^3=1/(2\pi^2)=2/(2\pi)^2$. No source consulted fixes either choice,
and nothing below depends on one: the normalisation volume $V$ cancels in any
ratio of spectra and is not fixed here.

The coefficients of Eq.~(\ref{eq:remainder}) follow from Eq.~(\ref{eq:dCl}) by
two expansions and one angular average, and are short enough to set out in full.

Change variables in the first term of Eq.~(\ref{eq:dCl}) to
$\bm L=\bm\ell-\bm\ell'$, so that
\begin{equation}
  \Delta C_\ell = \int\!\frac{\dd^2L}{(2\pi)^2}\,
    \bigl[(\bm\ell-\bm L)\!\cdot\!\bm L\bigr]^2 C^\psi_L\,C_{|\bm\ell-\bm L|}
    - \ell^2R\,C_\ell ,
  \label{eq:dCl-L}
\end{equation}
and write $\mu$ for the cosine of the angle between $\bm L$ and $\bm\ell$. The
kernel is exact,
\begin{equation}
  \bigl[(\bm\ell-\bm L)\!\cdot\!\bm L\bigr]^2
   = \ell^2L^2\mu^2 - 2\ell L^3\mu + L^4 ,
  \label{eq:kern}
\end{equation}
and the argument of the spectrum needs
\begin{equation}
  |\bm\ell-\bm L| - \ell
   = -L\mu + \frac{L^2}{2\ell}\bigl(1-\mu^2\bigr) + O(L^3/\ell^2) ,
  \label{eq:argexp}
\end{equation}
so that $C_{|\bm\ell-\bm L|}=C_\ell+(|\bm\ell-\bm L|-\ell)C'_\ell
+\frac12L^2\mu^2C''_\ell+\ldots$ Multiplying Eqs.~(\ref{eq:kern})
and~(\ref{eq:argexp}) and keeping every term through $L^4$, the average over the
direction of $\bm L$ uses only $\la\mu\ra=\la\mu^3\ra=0$, $\la\mu^2\ra=\frac12$
and $\la\mu^4\ra=\frac38$, and gives
\begin{equation}
  \frac{\ell^2L^2}{2}\,C_\ell
  + L^4\Bigl[\ts\frac{3}{16}\ell^2C''_\ell
   + \ts\frac{17}{16}\ell C'_\ell + C_\ell\Bigr] .
  \label{eq:angavg}
\end{equation}
The first term integrates to $\ell^2RC_\ell$ by the definition of $R$ and
cancels the subtraction in Eq.~(\ref{eq:dCl-L}) exactly, which is the
cancellation of Sec.~\ref{sec:quartic} seen term by term rather than argued from
the rigid-translation limit. What is left is Eq.~(\ref{eq:remainder}).

Where each coefficient comes from is recorded here, because
$\frac{17}{16}$ is the only one that is not a single term. The $C''$
coefficient comes from $\ell^2L^2\mu^2\times\frac12L^2\mu^2C''$ alone, and is
$\frac12\la\mu^4\ra=\frac3{16}$. The $C'$ coefficient has two sources:
$-2\ell L^3\mu\times(-L\mu C')$ gives $2\la\mu^2\ra=1$, and
$\ell^2L^2\mu^2\times\frac{L^2}{2\ell}(1-\mu^2)C'$ gives
$\frac12(\la\mu^2\ra-\la\mu^4\ra)=\frac1{16}$. The $C$ coefficient is
$L^4\times C_\ell$ and carries no average at all. Nothing in the three depends
on the shape of $C^\psi_L$, which is why Eq.~(\ref{eq:remainder}) factorises
into one integral over the lens spectrum and one bracket in $\ell$.

\section{The threading shift along a ray}
\label{app:shift}

Equation~(\ref{eq:dK}) is obtained in Sec.~\ref{sec:threading} from the frame
transformation of $A_a$, $\sigma_{ab}$ and $\omega_a$, Eq.~(\ref{eq:frame-transf}).
It can be had a second way, from the photon alone, and the two routes share no
step. This appendix gives the second. Its $Hv$ term is the one place the two
could differ, and it is worth seeing where that term comes from.

\emph{The pointwise map.} One null $p^a$ is split by two threadings. From
$\tilde E=-p^a\tilde u_a$ with $\tilde u^a=u^a+v^a$ and $v_au^a=0$,
\begin{equation}
  \tilde E = E\,(1 - v_ae^a) ,
  \label{eq:doppler}
\end{equation}
and from $\tilde e^a=p^a/\tilde E-\tilde u^a$, to first order,
\begin{equation}
  \tilde e^a = e^a - v^a + (v_be^b)\,(u^a+e^a) ,
  \label{eq:aber}
\end{equation}
which satisfies $\tilde e^a\tilde u_a=0$ and $\tilde e^a\tilde e_a=1$ as it must.
Projected into the rest space of $u^a$ this is
$h^a{}_b(\tilde e^b-e^b)=-v^a_\perp$, with
$v^a_\perp=v^a-(v_be^b)e^a$: aberration, as Sec.~\ref{sec:frameinv} states it.

\emph{The rate.} Each threading transports its own direction by
Eq.~(\ref{eq:de}),
\begin{equation}
  h^a{}_b\,p^c\nabla_c e^b = -E\,K^a_\perp ,
  \qquad
  \tilde h^a{}_b\,p^c\nabla_c\tilde e^b = -\tilde E\,\tilde K^a_\perp .
  \label{eq:tworates}
\end{equation}
Write $\tilde h^a{}_b = h^a{}_b + v^au_b + u^av_b$ and
$\tilde e^b = e^b + \delta^b$ with $\delta^b$ read off Eq.~(\ref{eq:aber}), and
contract the second of Eqs.~(\ref{eq:tworates}) with the screen projector of
$u^a$, $P^a{}_b \equiv h^a{}_b - e^ae_b$, which annihilates both $u^b$ and
$e^b$. Three terms survive at first order.

The first is $P^a{}_bh^b{}_cp^d\nabla_de^c=-EK^a_\perp$.

The second is $P^a{}_b\,p^c\nabla_c\delta^b$. The part of $\delta^b$ along
$u^b+e^b=p^b/E$ contributes nothing: $p^c\nabla_c(p^b/E)$ is again proportional
to $p^b$, and $P^a{}_b(u^b+e^b)=0$. What is left is
$-P^a{}_b\,p^c\nabla_cv^b = -E\,(\dd v/\dd t)^a_\perp$, with
$\dd/\dd t=u^c\nabla_c+e^c\D_c$ the derivative along the ray, since
$P^a{}_b\,e^c\nabla_cv^b = P^a{}_b\,e^c\D_cv^b$ exactly.

The third is $P^a{}_b(v^bu_c+u^bv_c)\,p^d\nabla_de^c$. The $u^b$ piece dies on
$P^a{}_b$; the other leaves $v^a_\perp\,(u_c\,p^d\nabla_de^c)$, and that
coefficient is the redshift itself. Differentiating $u_ce^c=0$ along the ray and
using Table~\ref{tab:conventions}, C2.2,
\begin{align}
  u_c\,p^d\nabla_de^c &= -e^c\,p^d\nabla_du_c
  \nonumber\\
  &= -E\bigl(\ts\frac13\Theta + A_ce^c + \sigma_{cd}e^ce^d\bigr)
  = \frac{1}{E}\frac{\dd E}{\dd\lambda} ,
  \label{eq:redshiftcoef}
\end{align}
by Eq.~(\ref{eq:dE}). It multiplies the first-order $v^a_\perp$, so only its
background value $-EH$ is kept, and the third term is $-EHv^a_\perp$.

Adding the three and cancelling $-E$,
\begin{equation}
  \tilde K^a_\perp - K^a_\perp
  = \Bigl(\frac{\dd v}{\dd t}\Bigr)^{\!a}_{\!\perp} + H\,v^a_\perp ,
  \label{eq:dK-again}
\end{equation}
which is Eq.~(\ref{eq:dK}).

Three things about the $Hv$. It comes from the third term and
from nowhere else, that is, from the first-order difference between the two
projectors and not from the boost: the two threadings carry their screens along
the ray differently. By Eq.~(\ref{eq:redshiftcoef}) its coefficient is the
redshift rate, which is the covariant form of the statement that a boost mixes
energy with direction, so the rate at which the energy changes feeds the
difference of the directions. And differentiating the pointwise map
Eq.~(\ref{eq:aber}) at a fixed projector would give $\dd v_\perp/\dd t$ alone
and miss it; that is an error, not a second answer.

\section{The tensor sector}
\label{app:tensor}

This appendix does no new calculation. It says which of the results above are
statements about the scalar sector and which are statements about the operator,
and it records the one place where the difference is important.

This split is useful because of what it carries directly from general
relativity. Relative to $u^a$ the free field splits into
$E_{ab}$ and $H_{ab}$, Eq.~(\ref{eq:weyl}). At first order the scalar sector
carries no magnetic part: $H_{ab}$ is sourced by the vector and tensor sectors,
and gravitational waves carry both parts together. That is the content of the
third consequence of Sec.~\ref{sec:weyl} --- $H_{ab}$ has no Newtonian analogue
--- read the other way round.

You might ask how this reaches the photons. That is the point: it reaches them
exactly as the scalar sector does. The tidal field never enters the hierarchy directly; it enters through the
shear, because the shear propagation equation carries $-E_{ab}$ as its source,
and it is $\sigma_{ab}$ that Table~\ref{tab:coefficients} couples to the
multipoles. Nothing in that table knows what sourced the shear. The tensor
contribution therefore appears in the same rows, with the same $\ell$-weights,
through the same $\sigma^a{}_be^b$ in Eq.~(\ref{eq:K}).

Both of the paper's claims are statements about the
operator and not about what sources it, so both carry over without extension.
The identification of Sec.~\ref{sec:content} is that the $O(\ep^2\ell)$ content
of $K_\perp^a\partial/\partial e^a$ is lensing and observer aberration; the
relabelling of Sec.~\ref{sec:relabel} is algebra on the multipoles, generated by
Eq.~(\ref{eq:dK}), which is a statement about two congruences and not about the
matter or the mode content. Lensing by tensors is the row it already is in
Table~\ref{tab:physics}, and the enumeration of Table~\ref{tab:enumeration}
reaches it because the count there is over the degrees of freedom of the map
rather than over what sources them.

One thing, and it is used above. \emph{The Newtonian threading is shear-free
for scalar modes only.} The congruence orthogonal to the
constant-time surfaces acquires a shear from the tensor sector, so with
gravitational waves present that threading does carry an $\ell\pm2$ shear
coupling. Every statement above that the Newtonian threading has
$\tilde\sigma_{ab}=0$ --- Secs.~\ref{sec:threading}, \ref{sec:lensing},
\ref{sec:curvhalf} and~\ref{sec:newtonian-pair} --- is a scalar-sector statement
and is used as one.

Section~\ref{sec:curvhalf} is the one where the vanishing does more than
simplify, and the limit there is sharper than a missing term.
Equation~(\ref{eq:conformalC}) writes the connection difference as the one a
conformal rescaling generates, which the sections of the Newtonian threading
admit because their scalar geometry is conformally flat at first order. A
transverse trace-free shear leaves an intrinsic geometry that no conformal
factor reproduces: in the tensor sector there is no $\Phi_C$, and
Eq.~(\ref{eq:PhiCPhiH}) has no tensor counterpart. Nor can one be written by
carrying the $-\frac13\Theta\sigma_{ab}$ of Eq.~(\ref{eq:gauss-tf}) across into
a scalar Hessian. The divergence forbids it. For any scalar $S$,
$\D^a\D_{\la a}\D_{b\ra}S=\frac23\D_b\D^2S$, while
$\D^a\sigma^{T}_{ab}=0$, so equating the two would force $\D_bS=0$ on every
mode of nonzero wavenumber.

What Eq.~(\ref{eq:gauss-tf}) does give, projected on the tensor sector, is
\begin{equation}
  \bigl({}^{(3)}\!R_{\la ab\ra}\bigr)^{T}
    = E^{T}_{ab} + \ts\frac12\pi^{T}_{ab} - H\,\sigma^{T}_{ab}
  \label{eq:gauss-tensor}
\end{equation}
at leading order in the tensor amplitude, and the intrinsic geometry it
describes needs a tensor connection perturbation that Eq.~(\ref{eq:conformalC})
cannot represent. The sections therefore carry a tensor contribution to the
deflection, and this paper does not compute it: the division into two halves in
Sec.~\ref{sec:curvhalf}, and with it their equality for a point mass, is a
scalar-sector result and is used as one.

The aberration side is untouched by any of that. It reaches the photons through
$\sigma^a{}_be^b$ in $K^a$ whatever sources the shear, which is the statement
Sec.~\ref{sec:content} and Table~\ref{tab:enumeration} use, and it is what the
frame-covariance argument depends on: the two threadings carry the same
operator, and they do whatever sources $K^a$. What the vanishing shear gives
elsewhere is only that the Newtonian division has two terms instead of three,
which is why it is the convenient one to compute in.

The amplitude is set by the tensor shear, and getting it requires the tensor
sector's own evolution, which is outside this paper; the covariant treatment of
that sector at linear order is Ref.~\cite{Challinor2000gw}. So is polarization,
which is where a treatment of the tensor contribution to the observed sky would
have to go next. The hierarchy used here carries the intensity alone;
Ref.~\cite{Challinor2000pol} extends the $1+3$ covariant formalism to the
polarization multipoles, in the same PSTF representation.

\section{The total bending and the point-mass limit}
\label{app:endpoints}

The total bending is $d=-\int\nabla_{\!\perp}(\Phi_A-\Phi_H)\,\dd\chi$, which
for $\Phi_H=-\Phi_A$ is $-2\int\nabla_{\!\perp}\Phi_A\,\dd\chi$, so
Eq.~(\ref{eq:bendK}) is exactly $d/2$; for a point mass, $2GM/b$ against
$4GM/b$. In the sign convention of Eq.~(\ref{eq:shear-source}) the rotation of
Eq.~(\ref{eq:bendH}) is attractive for $\Delta>0$. The unweighted $d$ is the
right comparison for the direction integrals,
Eqs.~(\ref{eq:bendK})--(\ref{eq:bendH}); the remapping of the sky is the
lens-weighted $\delta x_\perp$ of Eq.~(\ref{eq:deltax}), and $d$ enters only
the aberration at emission. The section term has the same angular operator as
the acceleration term, $(1-\mu^2)\partial_\mu$, so it adds to the first of
Eqs.~(\ref{eq:legendre}) with the same finite-difference structure, and the
cancellation of Sec.~\ref{sec:aligned} is unchanged.

\section{The enumeration, term by term}
\label{app:enumeration}

Table~\ref{tab:enumeration} is the enumeration Sec.~\ref{sec:residue}
summarises: every product of two $O(\ep)$ quantities the geodesic map produces,
the order it carries, and the term of Eq.~(\ref{eq:master}) it comes from. It
is placed here for the same reason as Table~\ref{tab:conventions}: it is a
reference to be returned to.

\begin{table*}[tp]
\caption{The enumeration. With the endpoints fixed, the observed brightness
depends on the emitted field, the null geodesic between the endpoints and
$u_R^a$, so at $O(\ep^2)$ every term is one $O(\ep)$ perturbation of the map or
of the observer's frame multiplied by one $O(\ep)$ field. The map's
perturbation at the emission point has six components --- two transverse and one
radial of position, two of direction and one of energy --- and the observer's
frame has three, and there is a row below for each. The order column is settled
by one question: whether a transverse or angular gradient acts on the high-$\ell$
pattern of $\tau$. The rows are forced by that count and not assembled --- it is a
count of the arguments the observed brightness has, not a list of effects that
have been thought of, and what lies outside it is named in
Sec.~\ref{sec:residue} --- which is
what entitles Sec.~\ref{sec:content} to say \emph{and nothing else}, and what
discharges the generic non-aligned kernel: the identification is an operator
statement, so no separate amplitude is required for a configuration in which the
wavevectors are not aligned. Rests-on entries are lines of
Table~\ref{tab:conventions}.}
\label{tab:enumeration}
\begin{ruledtabular}
\begin{tabular}{>{\raggedright\arraybackslash}p{0.145\textwidth}>{\raggedright\arraybackslash}p{0.175\textwidth}c>{\raggedright\arraybackslash}p{0.34\textwidth}l}
Perturbation & Acting on & Order & What it is, and why & Rests on \\
\colrule
$\delta x_\perp$ \ (2)
  & $\nabla_{\!\perp}\tau_E$, the emitted pattern
  & $\ep^2\ell$
  & \emph{Lensing} of the primary, Eq.~(\ref{eq:deltax}); the efficiency
    $(\chi_*-\chi)/\chi_*$ comes from the distance factor of Eq.~(\ref{eq:dtau})
  & C2.5 \\
$\delta x_\perp$ \ (again)
  & the linear secondaries along the ray
  & $\ep^2\ell$
  & \emph{Lensing} of the secondaries, each with its own efficiency
    $(\chi'-\chi)/\chi'$ for a source at $\chi'$
  & C2.5 \\
$\delta x_\parallel$ \ (1)
  & $\partial\tau_E/\partial\chi$
  & $\ep^2\ell^0$
  & Time delay and the perturbed last-scattering surface. The gradient is
    radial, so no factor of $\ell$: the modes dominating at high $\ell$ are
    nearly transverse, so the radial component does not grow with it. About
    $0.1\%$ in $C_\ell^{TE}$ at $\ell\sim10^3$ \cite{HuCooray2001}. Part of the
    geodesic map, hence frame-invariant
  & C1.5 \\
$\delta e_\perp|_E$ \ (2)
  & $\partial_e\tau_E$, the intrinsic direction dependence
  & $\ep^2\ell^0$
  & Aberration at emission. The operator carries $\ell$, but the field it acts
    on has $\ell\le2$ in the baryon frame, so the coupling is $O(1)$, and
    beyond the dipole it is suppressed by tight coupling
    \cite[footnote~28]{ZibinScott2008}
  & C2.2, C6.1 \\
$\delta E/E$ \ (1)
  & $\tau^{(1)}$, and $\bar\tau$ through its energy dependence
  & $\ep^2\ell^0$
  & Rees--Sciama, including the moving lens. The gradients act on $\Phi$ and
    not on $\tau$, so the standard scaling argument applies unchanged
  & C2.2 \\
$v_{R\perp}$ \ (2)
  & $\partial_e\tau$ at $R$
  & $\ep^2\ell$
  & \emph{Observer aberration}: the one $O(\ep^2\ell)$ term fixed by $u_R^a$
    rather than by the geometry between the endpoints
  & C1.5, C6.1 \\
$v_R\!\cdot\!e$ \ (1)
  & $\tau$ at $R$
  & $\ep^2\ell^0$
  & Observer modulation at $R$. No gradient acts on $\tau$ at all
  & C1.5 \\
\colrule
any of the nine
  & $\bar\tau$
  & ---
  & \emph{Zero.} The background is homogeneous and isotropic, so
    $\nabla_{\!\perp}\bar\tau=0$ and $\partial_e\bar\tau=0$; only its energy
    dependence survives, in the $\delta E/E$ row
  & C5.1 \\
the source of $K^a$
  & ---
  & ---
  & \emph{Not a further row.} Whatever sources $K^a=A^a+\sigma^a{}_be^b
    +\ep^a{}_{bc}e^b\omega^c$, its scalar, vector and tensor contributions all
    enter through $\delta x_\perp$ and $\delta e_\perp$, counted above;
    $\omega^a$ vanishes at first order for scalar modes, the tensor shear does
    not, and is the lensing by tensors of Table~\ref{tab:physics}
  & C2.2 \\
three $O(\ep)$ factors
  & ---
  & $\ep^3$
  & Beyond this order: post-Born corrections, and the curl of the deflection
    for scalar modes --- a tensor shear curls it at Born order, which is the
    row above
  & C0.1 \\
\end{tabular}
\end{ruledtabular}
\end{table*}

\end{document}